\documentclass[]{aa}
\usepackage{natbib}
\usepackage{longtable,lscape}
\usepackage{amsmath}
\bibpunct{(}{)}{;}{a}{}{,}
\usepackage{graphicx}

\usepackage[colorlinks=true,citecolor=blue]{hyperref}
\usepackage{color}
\usepackage{xspace}
\usepackage{dcolumn}
\usepackage{longtable}
\usepackage{lscape}
\usepackage{wasysym}
\usepackage{rotating}
\usepackage{afterpage}
\usepackage{mathtools}

\newcommand{\MJup}{M$_{\mathrm{Jup}}$\xspace}

\newcommand{\MSun}{M$_{\odot}$\xspace}

\newcommand{\mic}{$\mu$m\xspace}
\newcommand{\as}{\hbox{$^{\prime\prime}$}\xspace}

\usepackage[varg]{txfonts}
\begin{document}

\title{Hints of planet formation in progress around DR\,Tau from observations with LBT SHARK-NIR and SHARK-VIS} 

\author{ D.~Mesa\inst{1}\thanks{Corresponding author: \email{dino.mesa@inaf.it}},
G.~Li~Causi\inst{2,3}, R.~Gratton\inst{1}, E.~Carolo\inst{1}, S.~Antoniucci\inst{2},
J.~Farinato\inst{1}, L.~Marafatto\inst{1}, E.~Rigliaco\inst{1}, 
D.~Barbato\inst{1}, F.~Battaini\inst{1}, M.~Bergomi\inst{1}, A.~Bianco\inst{4}, P.~Cerpelloni\inst{1}, 
F.~D'Alessio\inst{2}, S.~Desidera\inst{1},
S.~Di~Filippo\inst{1}, V.~D'Orazi\inst{5,2}, D.~Fasano\inst{2},
T.S.~Gomes~Machado\inst{1}, D.~Greggio\inst{1}, P.~Grenz\inst{6}, F.~Laudisio\inst{1},
C.~Lazzoni\inst{1}, J.~Leisenring\inst{6},  A.~Lorenzetto\inst{1}, E.~Marini\inst{2}, L.~Merli\inst{1}, L.~Mohr\inst{7}, 
F.~Pedichini\inst{2,3},
R.~Piazzesi\inst{2}, D.~Ricci\inst{1}, K.\,K.\,R.~Santhakumari\inst{1}, R.~Simonetto\inst{1,8}, V.~Testa\inst{2}, 
G.~Umbriaco$^{9,1}$, P.~Vaccari\inst{2}, D.~Vassallo\inst{1}, M.C.~Vega~Pallauta\inst{7,10}, V.~Viotto$\inst{1}$, A.~Zurlo\inst{11,12}
}

\institute{
\inst{1}INAF - Osservatorio Astronomico di Padova, Vicolo dell'Osservatorio 5, 35122 Padova, Italy \\
\inst{2}INAF - Osservatorio Astronomico di Roma, Via Frascati 33, 00078, Monte Porzio Catone, Italy \\
\inst{3}ADONI - Istituto Nazionale Italiano per l'Ottica Adattiva \\
\inst{4}INAF - Osservatorio Astronomico di Brera, via E. Bianchi 46, 23807 Merate, Italy \\
\inst{5}Dipartimento di Fisica, Università di Roma Tor Vergata, via della Ricerca Scientifica 1, 00133 Rome, Italy \\
\inst{6}Department of Astronomy and Steward Observatory, The University of Arizona, 933 North Cherry Ave, Tucson, AZ 85721, USA\\
\inst{7}Max-Planck-Institut f{\"u}r Astronomie, K{\"o}nigstuhl 17, 69117 Heidelberg, Germany \\
\inst{8}Dipartimento di Fisica e Astronomia, Università degli Studi di Padova, Vicolo dell’Osservatorio 3, 35122 Padova, Italy \\
\inst{9}Dipartimento di Fisica e Astronomia ”Augusto Righi” - Alma Mater Studiorum Università di Bologna, via Piero Gobetti 93/2, 40129, Bologna, Italy\\
\inst{10}Fakult{\"a}t f{\"u}r Physik und Astronomie, Universit{\"a}t Heidelberg, Im Neuenheimer Feld 226, 69120 Heidelberg, Germany \\
\inst{11}Instituto de Estudios Astrofísicos, Facultad de Ingeniería y Ciencias, Universidad Diego Portales, Av. Ej\'ercito Libertador 441, Santiago, Chile \\
\inst{12}Millennium Nucleus on Young Exoplanets and their Moons (YEMS), Chile \\
}

   \date{Received 29 April 2026 / accepted dd October 2026}

\abstract
    {Protoplanetary disks are the primary sites of planet formation. However, the direct imaging of protoplanets still embedded within their parental disks remains a significant challenge, with only a small number of confirmed detections to date and several candidates awaiting verification.}
    {We search for embedded planets in the protoplanetary disk of the young star DR\,Tau by targeting accretion signatures from forming sub-stellar objects. Using narrow-band imaging, we observe both visible $H\alpha$ and near-infrared Paschen-$\beta$ ($Pa\beta$) emission to identify actively accreting companions.
    }
    {To this end, we performed high-contrast observations of DR\,Tau using the SHARK-VIS (0.40–0.95~\mic) and SHARK-NIR (0.96–1.70~\mic) imagers at the LBT. The study uses a combination of narrow-band filters to isolate accretion signatures and broad-band filters for multi-wavelength characterization in both the visible and near-infrared.
    }
    {Our observations recover the primary disk structures previously identified in the literature, albeit at low signal-to-noise (S/N). We identify two compact features with S/N between 6 and 13 in both narrow- and broad-band filters. The first, an elongated structure, matches previous SPHERE detections, while the second, more compact feature, is now confirmed despite its low S/N in earlier datasets. A comparison over a seven-year baseline confirms the stability of both structures. Their observed orbital motion is consistent with Keplerian rotation, potentially suggesting the presence of an undetected planetary-mass companion ($\sim$few $M_{\text{Jup}}$) at the location of the second feature. However, our analysis does not confirm H$\alpha$ or Pa$\beta$ emission originating from localized accretion. Instead, we conclude that the observed flux is likely stellar light reflected by dust at these locations and discuss alternative mechanisms that could account for such detections.
    }
{}
   

\titlerunning{DR\,Tau}
\authorrunning{Mesa et al.}
   \maketitle
%

\section{Introduction}
\label{intro}

Protoplanetary disks surrounding stars with ages up to a few Myr are considered the natal environment for planets \citep[see e.g.,][]{2012ApJ...756..133C,2014A&A...565A..15M}. Observations in the visible, in the near-infrared (NIR), in the mid-infrared (MIR) and at millimeter wavelengths have allowed the detection of different types of sub-structures in the protoplanetary disks including gaps and rings \citep[e.g.,][]{2017A&A...601A...7F,2018A&A...610A..24F,2018ApJ...869L..49I,2022ApJ...941..172L,2024A&A...682A..55M,2025ApJ...989...69L}, cavities \citep[e.g.,][]{2017AJ....154...33A,2017A&A...607A..55V,2017A&A...597A..32V,2021MNRAS.502.5779N}, and spirals \citep[e.g.,][]{2020A&A...637L...5B,2021A&A...654A..35B,2022A&A...658A..63M,2025A&A...699A.237G}. More than one single mechanism can be implied in the formation and the evolution of such structures and, moreover, structures in different disks
can be explained with different mechanism. However, one of the most accepted scenario to explain such structures is that they are caused by the interactions between the disk and unseen companions embedded in the disk itself. However, up to now, only in a few cases (PDS \, 70: \citealt{2018A&A...617A..44K,2019NatAs...3..749H}; 2MASS J16120668-3010270:  \citealt{Li2025}; AB\,Aur: \citealt{2022NatAs...6..751C,2025ApJ...990L..42C, Bowler2025}; WISPIT-2:  \citealt{2025ApJ...990L...8V, Close2025,Lawlor2026}) has it been possible to directly image planetary mass companions embedded in protoplanetary disks. Additional cases, such as HD\,169142 \citep{2017ApJ...850...52P,2018MNRAS.473.1774L,2019A&A...623A.140G,Hammond2023}, still need solid confirmation. The direct detection of the seeds of forming planets has been proposed in a massive, spiral-armed disk around the FUor object V960 Mon \citep{2023ApJ...952L..17W}, together with a young planet just formed via gravitational instability around the same star \citep{2025ApJ...988L..30D}. 

In this work, we present new observations of the DR\,Tau system at both visible and NIR wavelengths using the two new imagers installed at the Large Binocular Telescope (LBT), SHARK-VIS \citep{pedichini22,pedichini24} and SHARK-NIR \citep{2015IJAsB..14..365F,2022SPIE12185E..22F,2022SPIE12184E..3VM}. The paper is organized as follows. In Sec.~\ref{s:target} we describe the target of the observations, presenting the current knowledge about it. In Sec.~\ref{s:obs} we describe our observations and the corresponding data reduction procedure. In Sec.~\ref{s:res} we present the main results of our observations, which are then discussed in Sec.~\ref{s:dis}. Finally, in Sec.~\ref{s:conclusion} we give our conclusions.


\section{Target description}
\label{s:target}

DR\,Tau is a classical T Tauri star located in the Taurus-Auriga star-forming region, at the south-eastern edge of Lynds 1558  \citep{2022A&A...658A..63M}. Its measured distance from the Sun is 192.97$\pm$1.24~pc \citep{2021A&A...649A...1G}, larger than the distance of the main region ($\sim$140~pc). It is known to exhibit strong photometric and spectral variability \citep{2001AJ....122.3335A,2007A&A...461..183G} so that it is classified as EXOr variable \citep[e.g.,][]{2009ApJ...693.1056L} and a high mass accretion rate \citep[$4.8\times10^{-7}$ M$_{\odot}$ yr$^{-1}$;][]{2019A&A...632A..32M}. This causes a high and variable veiling that does not allow for precise spectral classification of the star, that ranges between K4 and M0 \citep{2011A&A...535A...6P,2019ApJ...882...49L,2019A&A...632A..32M,2022A&A...667A.124G}. The same uncertainties are present in the mass determination. However, \citet{2021ApJ...908...46B} found for this star a mass of $1.18^{+0.59}_{-0.44}$~\MSun exploiting measurements of the Keplerian rotation of the disk. This agrees well with the value of $1.12\pm 0.10$ \MSun obtained by the same \citet{2021ApJ...908...46B} using the magnetic models by \citet{Feiden2016} and indicates a slightly super-solar mass. This value for the mass is partially in contrast with the spectral type determinations described above. However, this discrepancy could be mainly due to the veiling described above that impact clearly on the spectral type but not on the methods described by \citet{2021ApJ...908...46B}. For this reason, we decided to adopt this value for the mass in the following of this work. The age estimate for the system ranges between 0.9~Myr \citep{2019A&A...632A..32M} and 3.2~Myr \citep{2019ApJ...882...49L}. However, a recent estimate of the system age places it at the younger part of the age range excluding ages larger than $\sim$ 2~Myr and giving the higher probability for ages around 1~Myr \citep{2026A&A...715A..21G}.

Disk emission has been widely studied, allowing the detection of the signature of different molecules such as H$_2$O \citep{2013ApJ...770L..14B,2024A&A...689A.330T}, CO \citep{2019ApJ...882...49L}, SO, DCO$^+$ and and H$_2$CO \citep{2023ApJ...943..107H}. The dust disk was resolved at a wavelength of 1.3~mm by \citet{2019ApJ...882...49L} finding a radius of 51~au, an inclination of $5.4^{\circ +2.1}_{-2.6}$ and a position angle of $3.4^{\circ +8.2}_{-8.0}$ while a much larger radius of 246~au was found for the gas by \citet{2021ApJ...908...46B} exploiting the $^{13}$CO and C$^{18}$O molecular lines. Recently, \citet{2024A&A...686A.117T} 
and \citet{2024A&A...689A.330T} proposed a misalignment between the inner and the outer disk with an higher inclination for the inner disk ($10^{\circ}$--$23^{\circ}$) to explain the emission of the CO and of the H$_2$O molecules using JWST-MIRI
data.

Observing DR\,Tau in scattered light with VLT/SPHERE, \citet{2022A&A...658A..63M} identified two previously unknown spiral arms. They suggested that a late infall of material causes the southern one, while the north-eastern one could be related to a planet in formation. They indeed detected an elongated structure at about 0.3\as from the star that they suggested could be a hint of a planet still embedded in its dust envelope and that could launch the north-eastern spiral arm.
However, \citet{2023ApJ...943..107H}, using millimeter interferometry observations with NOEMA did not confirm such detection that then still need to be confirmed.

\section{Observations and data reduction}
\label{s:obs}

\subsection{SHARK-NIR data}
\label{s:sharknir}

\begin{table*}[!htp]
  \caption{Characteristics of each DR\,Tau observation. In the Median DIMM seeing column the two values in parenthesis 
  represents the minimum and the maximum value for the DIMM seeing during the observation. In the seventh column DIT
  stays for detector integration time.} \label{t:observation}
\centering
\begin{tabular}{ccccccccc}
\hline\hline
Instrument & Obs. date  & Coronagraph &  Filter  & Median DIMM seeing  & Field rotation & DIT   & Total exposure \\
           &            &             &          &         (\as)       &   ($^{\circ}$) &   (s) &     (s)        \\
\hline
SHARK-NIR  & 2025 November 01 &  Gauss & $NB\_PaB$  & 1.14 (0.98/1.45) &  65.0   &   90.9   & 6093.9 \\
SHARK-NIR  & 2025 November 02 &  Gauss & $BB\_H$    & 0.81 (0.67/1.07) &  83.9   &   61.5   & 7998.5 \\
SHARK-VIS  & 2025 November 02 & No coro&    $H\alpha$    &  0.75 (0.5/1)   &  96.0  &   5.0       &  10800.0     \\
SHARK-VIS  & 2025 November 02 & No coro&    $Cont$    &  0.75 (0.5/1)    &  96.0  &   5.0       &  10800.0     \\
\hline
\end{tabular}
\end{table*}

We observed DR\,Tau during two consecutive nights (November, 1 and 2 2025 UT) with the LBT SHARK-NIR instrument
designed to exploit the LBT Adaptive Optics (AO) system called SOUL \citep{2016SPIE.9909E..3VP,2023aoel.confE..80P}. During the first night we adopted the {\it Paschen-$\beta$} ({\it Pa$\beta$}) narrow band filter ({\it NB\_PaB} with $\lambda_c$=1.282~\mic and $\Delta\lambda$=0.023~\mic) while during the second night we observed using the broad-band {\it H} filter ({\it BB\_H} with $\lambda_c$=1.6~\mic and $\Delta\lambda$=0.218~\mic). The field-of-view of SHARK-NIR has a radius of 8\as while the pixel scale is of 14.5 mas\,pix$^{-1}$. The observations were performed in pupil stabilized mode. The rotation of the field of view allowed us to implement the angular differential imaging \citep[ADI;][]{2006ApJ...641..556M} method to subtract speckle noise. The main characteristics of each observation are listed in Table~\ref{t:observation}. Before and after each series of coronagraphic observations, we also took images of the stellar point-spread function (PSF) outside the coronagraph with the aim of estimating the final contrast reached with our observations and to estimate the contrast of any possible detected object. To avoid saturation, we performed these observations by introducing in the optical path a neutral density filter, namely the {\it ND2} filter for the {\it Pa$\beta$} observation and the {\it ND3} filter for the broad-band~{\it H} observation.

These data were pre-processed with the aim of reducing detector noise \citep[see][for more details on the procedure]{2025MNRAS.536.1455M} using a dedicated {\tt Python} procedure that performs reference pixel correction, bias subtraction, and linearity correction. Then it collapses the ramps (set of reads taken at uniform time intervals) of which any individual exposure is composed into single frames using an up-the-ramp sampling algorithm. The resulting images were then reduced using a different {\tt Python} pipeline prepared appropriately for the SHARK-NIR data. The first step of the pipeline consists of subtracting the dark and dividing by flat-field images. We then fine-centered each frame by utilizing the four symmetrical spots created by the SHARK-NIR internal deformable mirror. The procedure fits two lines passing through two opposite spots and considers their intersection as the position of the center of the PSF. We also exploits such spots to automatically discard those where they were not visible. We also couple this to a visual check to discard additional frames that were compromised like, for example, those frames where the star was not adequately masked by the coronagraph because of the residual jitter. Following this procedure we remained with 43 frames out of 67 for the {\it Pa$\beta$} case and with 128 frames out 130 from the broad-band $H$ datasets thanks to the much better conditions in the second night. Finally, we applied a post-processing procedure based on the ADI method and using the principal components analysis 
\citep[PCA;][]{2012MNRAS.427..948A,2012ApJ...755L..28S}. We also implemented a version of the iterative
PCA-ADI \citep[PCA-IADI;][]{2022A&A...668A..50S} prepared for SHARK-VIS (see Sec.~\ref{s:sharkvis} for details) and adapted to the SHARK-NIR case but we found that it was not able to improve the final contrast in this case, probably because of the
much lower number of frames of which the SHARK-NIR dataset is composed. We will then present only the PCA results for the 
SHARK-NIR case.

\subsection{SHARK-VIS data}
\label{s:sharkvis}

In parallel with the SHARK-NIR observations described above we also observed with SHARK-VIS \citep{pedichini22,pedichini24}  using the prismatron dual-band splitter, which provides simultaneous imaging in two narrow-band {\it H$\alpha$} filters both centered at 656.3 nm, one with a 1 nm bandwidth (hereafter simply called ``{\it H$\alpha$}'') and the other with a 10 nm bandwidth (hereafter called ``Continuum''). Like for the SHARK-NIR case, also the SHARK-VIS observation exploits the SOUL AO system. The field stop was 2\farcs4 in diameter, and the plate scale was 6.43~mas\,pix$^{-1}$. No coronagraph was used as the unmasked stellar PSF is needed to align the frames, because SHARK-VIS does not have an internal deformable mirror to create the four symmetrical spots as SHARK-NIR does. Moreover, operating without a coronagraph enables investigating the region very close to the star to potentially exploit the enhanced spatial resolution at visible wavelengths. Furthermore in the visible we reach a maximum Strehl ratio of 30\% (only in the very best seeing conditions of about 0\farcs4 and with short integration times of tens of ms) and in these conditions the wavefront is far from being flat. As a result, including a coronagraph would simply result in dimming the central region as a mask without decreasing the PSF halo. Poor seeing conditions on November, 1 (1\as to 1\farcs5) yielded much worse results with respect to November, 2 (0\farcs5 to 1\as), so we only refer to the latter in this work. The main characteristic of these observations are listed in Table~\ref{t:observation}.

We processed the frames with the custom {\tt DReAMS} (Data Reduction, Analysis, and Modeling for SHARK-VIS) pipeline (Li Causi et al., in prep). We cleaned the frames for detector signatures through various steps. The first one consists in the subtraction of a cleaned average dark image that was separately acquired at the end of the observing night. We then performed a correction of the frame-by-frame variable bias pattern, mainly consisting of non-static column-by-column (and less intense row-by-row) bias differences, which we measured on each frame as the column and row averages, respectively, of the pixels outside the field stop. The third step consists in masking of cosmic rays (CRs), which we found as follows: first, we computed a candidate CR map via a 15-pixels wide box sigma filter (at 4.5 sigma) on the maximum-pixel frame stack after subtraction of its radial profile; then we refined this map via a temporal sigma filter (at 4.5 sigma) of the value of each candidate pixel from 10 frames before to 10 frames after respect to the frame where that pixel's value was maximum. Pixels flagged as CR were replaced with NaN for the rest of the processing, and finally replaced with zero before final ADI processing. Finally, we divided by the flat field map, previously acquired with the SHARK-VIS internal calibration unit.

We then performed an analysis of the PSF core on the cleaned frames via a Gaussian fit, to both compute the
peak location and select good frames, thanks to a manual thresholding on a plot of the peak properties. Namely, in this case 
we rejected all peaks with FWHM $<$ 8 pixels, with very low peak amplitude (when the AO loop opened), and with not converging 
Gaussian fit. The selected frames were then co-aligned to a fractional pixel, via Fourier translation, based on the measured 
peak locations. Finally, we searched for possible frames with an anomalous PSF (e.g., double-peaked or motion-blurred) and 
removed them (we found those frames thanks to an iterative process that finds the constant areas in a frame-index map 
indicating the frame where each pixel had its maximum value).
 After this cleaning, we ended up with 2117 selected frames (98\% of the raw ones), for a total of about 2h 56m of total exposure time. The quality of the PSF in both {\it H$\alpha$} and {\it Continuum} average stacks is good, with a full width at half maximum (FWHM) of 33 mas (5.16 pix, that is 1.95 $\lambda$\,D$^{-1}$), corresponding to a Strehl ratio of 15\%, and a minimum radial profile within the AO control radius of $2.5 \times 10^{-3}$ at 200 mas.

Finally, we performed a PCA-IADI method, in which the resulting structures
of the usual PCA-ADI process are recursively subtracted from the original frames to achieve a PSF model not affected by the 
circumstellar structures, thereby substantially compensating brightness attenuation and drastically reducing the 
self-subtraction artifacts with respect to the ADI results.

Since there can still be self-subtraction that is hard to get rid of, as described in the literature 
\citep[e.g,][]{2022A&A...668A..50S}, we tried different threshold values, for selecting the structures to subtract in
the IADI process, and different iteration schemes. Among them we choose to adopt a zero threshold (i.e., all positive pixels 
after each iteration are considered true objects in the next iteration) and to increase the number of PCA components by one at 
each iteration, starting from the first component and repeating for 100 iterations, which provided the final stack shown in 
Sec.~\ref{s:res}.


\section{Results}
\label{s:res}

\begin{figure*}[!htp]
\centering
\sidecaption
\includegraphics[width=12cm]{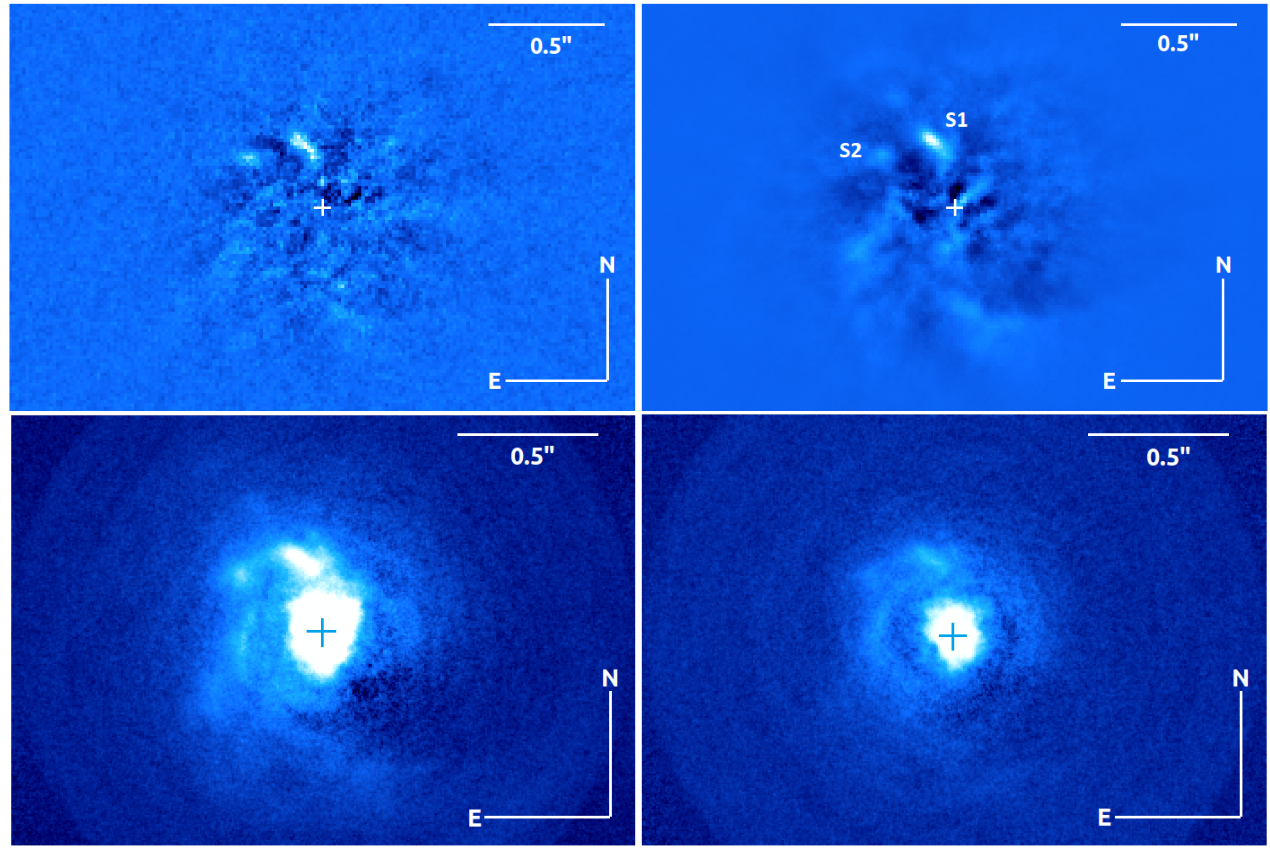}
\caption{Final images obtained both for SHARK-NIR (top panels) and SHARK-VIS (bottom panels). In the top left panel we display 
the final image obtained using the the narrow-band {\it Pa$\beta$} filter while in the top right panel we display the final 
image obtained using the broad-band~$H$ filter. Both the SHARK-NIR images have been obtained by applying a PCA-ADI method, 
subtracting five principal components. The white crosses represent the positions of the star in both images. In the bottom 
left panel we display the final image obtained using the {\it H$\alpha$} narrow-band filter of SHARK-VIS while in the bottom 
right panel we display the final image obtained using the ``$continuum$'' filter of SHARK-VIS. The same structures visible in 
the SHARK-NIR images are also visible in these images. The blue crosses represents the position of the star.}
\label{f:DRTau_finimg}
\end{figure*}

We display the final image obtained with the {\it Pa$\beta$} SHARK-NIR data on the top left panel of Fig.~\ref{f:DRTau_finimg} while the analogous result for the broad-band {\it H} data is shown in the top right panel. In the bottom left panel of Fig.~\ref{f:DRTau_finimg} we display the final image obtained from the SHARK-VIS {\it H$\alpha$} data while on the bottom right panel we display the final image obtained using the ``{\it continuum}'' data.

\subsection{Compact structures}
\label{s:compstruct}

\begin{figure}
\centering
\includegraphics[width=0.30\textwidth]{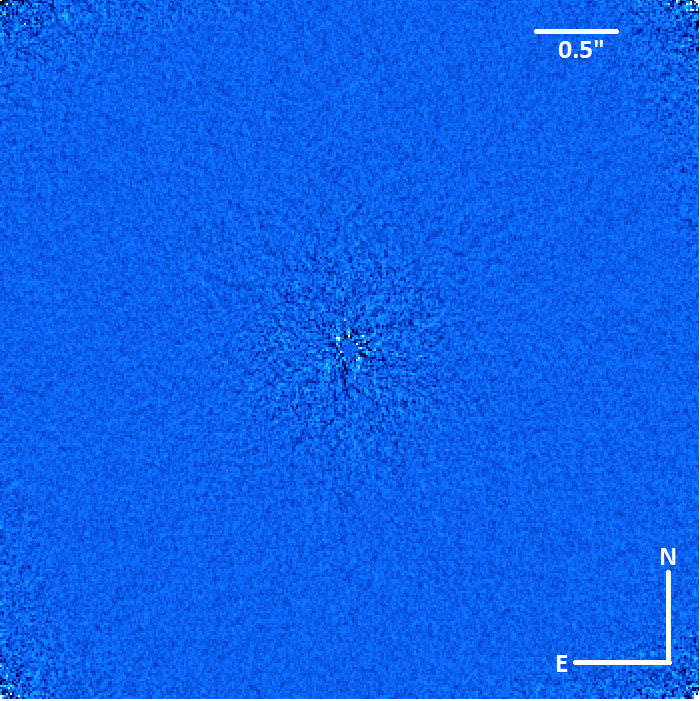}
\caption{Final image obtained from the subtraction of the ``$continuum$'' from the $H\alpha$ data with the aim to highlight 
possible emission caused by accretion. It is clear that we were not able to recover any structure from the disk of DR\,Tau.}
\label{f:subtractionVIS}
\end{figure}

For both the SHARK-NIR scientific filters, two structures are evident in the final images. We have labeled them as S1 and S2 in the top right panel of Fig.~\ref{f:DRTau_finimg} referred to the broad-band $H$ case where they are detected at larger signal-to-noise (S/N). Indeed, in narrow-band {\it Pa$\beta$} the S/N of the S1 structure is $\sim$6 while the S/N for the S2 structure is greater than 8. On the other hand, in the broad-band $H$ data, the S/N of the S1 structure is larger than 13, while that of the  S2 structure is around 10.6. The S1 structure is clearly elongated and it is just North of the star at a separation of 304$\pm$10~mas and position angle (PA) of 8.6$^\circ$$\pm$1.5$^{\circ}$ while the more compact S2 structure has a separation of 384$\pm$7~mas and a PA of 54.2$^{\circ}$$\pm$0.7$^{\circ}$. The astrometry of these two structures given above was estimated using the negative companion method \citep{2011A&A...528L..15B,2014A&A...572A..85Z}. The method consists in including a negative companion at the location of the detected source in such a way to cancel out its flux in the original dataset and reducing the modified dataset with the same high-contrast method used for the science data. The best estimation for the position of the companion is the one that minimizes the residuals in a small region around the position. We stress here that this method is optimized for point-like sources but it is, nonetheless, able to give us a good approximation of the correct position of the flux peak. \par

Adopting again the negative planet method and with the same caveats given for the astrometric case, we have calculated the contrast of these two structures. In the case of the {\it Pa$\beta$} data, S1 has a contrast of $(1.17\pm0.25)\times10^{-4}$ and for S2 we found a contrast of $(7.49\pm2.65)\times10^{-5}$. From the broad-band $H$ data we found contrasts of $(1.42\pm0.09)\times10^{-4}$ and  $(4.75\pm1.90)\times10^{-5}$ for the S1 and S2 structures, respectively. \par

The same structures are also visible in the SHARK-VIS final images with both the scientific filters. It is however apparent 
that the ``$continuum$'' data are much more noisy than the {\it H$\alpha$} data. This is mainly because the stellar flux (and 
therefore the flux reflected-off from the disk) in {\it H$\alpha$} line is higher due to the excess line emission from
accretion, which unbalance the 1/10 transmission ratio of the ``$continuum$'' channel to the {\it H$\alpha$} channel in the prismatron dual-band splitter. For this reason, we will refer to the {\it H$\alpha$} data in the following. Differently from the SHARK-NIR data, it is apparent in the SHARK-VIS data the presence of material, perhaps a section of the spiral arm, linking S1 with S2. Following the same methods devised for SHARK-NIR, we found a S/N of 7.13 for S1 and of 7.96 for S2. The S1 structure has a separation of 272$\pm$15~mas and a PA of 9.3$^{\circ}$$\pm$1.5$^{\circ}$ while the S2 structure has a separation of 362$\pm$8~mas and a PA of 55.9$^{\circ}$$\pm$0.7$^{\circ}$. We also calculated the contrast for the two structures finding a value of $7.90\times 10^{-5}$ for S1 and of $4.63\times 10^{-5}$ for S2. \par
The ``$continuum$'' filter in SHARK-VIS is aimed to be subtracted from the $H\alpha$ data to highlight the possible 
presence of emission mainly from accretion. This was indeed done also in this case but, as can be seen in 
Fig.~\ref{f:subtractionVIS} where we show the difference obtained subtracting the ``$continuum$'' image obtained using the 
PCA-ADI using 100 principal components from the $H\alpha$ image obtained in the same way, we were not able to recover any of the 
structures visible in the DR\,Tau disk. We note that to take into account the different flux in $H\alpha$ and in ``$continuum$'' 
data we have to scale the inner 50~mas of the latter image by a factor 3.5 before to perform the subtraction described above. 
Additionally, we note that because of the high noise present in the ``$continuum$'' image, this procedure quadratically adds this noise to the $H\alpha$ signal making the results less significative.

\subsection{Disk}
\label{s:disk}

\begin{figure*}[!htp]
\centering
\sidecaption
\includegraphics[width=12cm]{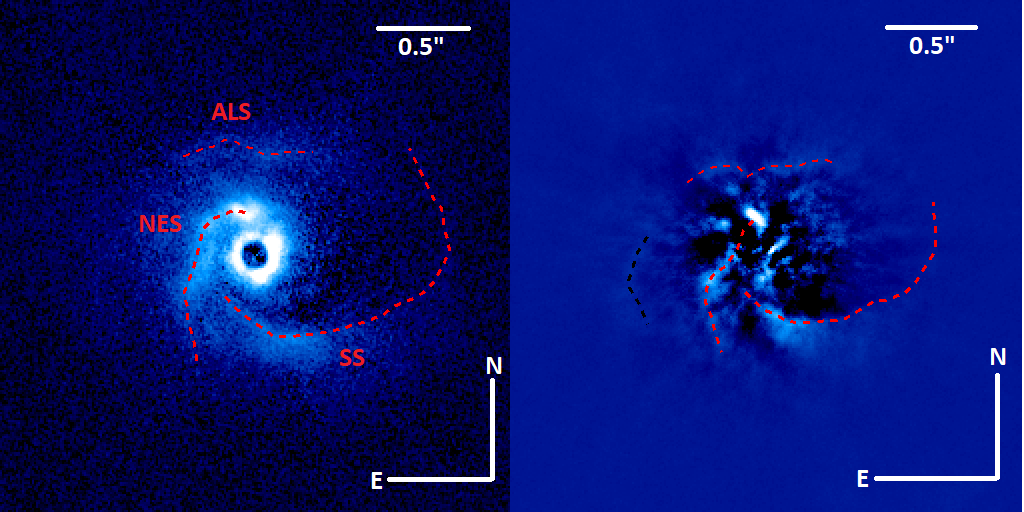}
\caption{Comparison of the SPHERE and SHARK results for the disk spirals around DR\,Tau. {\it Left panel:} Final image for the SPHERE polarized data from \citet{2022A&A...658A..63M}. Following that paper we  superimposed red dashed lines to highlight the main disk structures and we have tagged them with SS standing for southern spiral, NES standing for north-eastern spiral, and ALS standing for arc-like structure. {\it Right panel: } Final image from the SHARK-NIR broad-band H data with superimposed dashed lines on the same structures of the SPHERE case to highlight that the same disk structures have been recovered. The dark dashed line on the left represents 
a possible arm-like structure that was not visible in the SPHERE data.}
\label{f:diskall}
\end{figure*}

The main objective of these observations was the detection of compact structures, and they were not optimized for the detection of the disk. Indeed, they were not performed in polarized light as this observing mode is not available for the SHARKs and  this, in association with self-subtraction, has the consequence that the disk around DR\,Tau is barely visible. However, we are able to image the main disk structures observed by \citet{2022A&A...658A..63M}. To highlight this, in the left panel of Fig.~\ref{f:diskall} we display a SPHERE polarized data image adapted from \citet{2022A&A...658A..63M}. In this image we draw red dashed lines to highlight the three main disk structures that are the southern spiral, tagged with SS, the north-eastern arm, tagged with NES, and the arc-like structure tagged with ALS. In the right panel of Fig.~\ref{f:diskall} we display as comparison the final image from the SHARK-NIR broad-band $H$ data. Also on this image we draw red dashed lines to highlight the main disk structures. All the features in the SPHERE data are visible also in this case. In particular, SS is visible in all its extension while NES is much less visible, probably because it is suffering of a larger self-subtraction due to the presence of more disk material in its proximity, even if its presence is deducible in our data. Finally, we are able to recover the ALS structure just north of the star. Furthermore east from the star is also visible an extended structure that was not visible in the SPHERE data that is highlighted with a dashed dark line in the right panel of Fig.~\ref{f:diskall}. It is however unclear, given its very low S/N, if it is a real structure or a feature created by the data reduction. The S/N of the disk structures detection is generally low, and this prevents us from being able to calculate the main spiral characteristics, as done by \citet{2022A&A...658A..63M}. Moreover, the spirals and the arc structures present some clear deformation (more visible looking at the red dashed lines on the right panel of Figure~\ref{f:diskall}) that are more likely due to the low S/N of the detections than to the real characteristics of such structures.

\subsection{Contrast limits}
\label{s:contrast}

\begin{figure*}[!htp]
\centering
\sidecaption
\includegraphics[width=12cm]{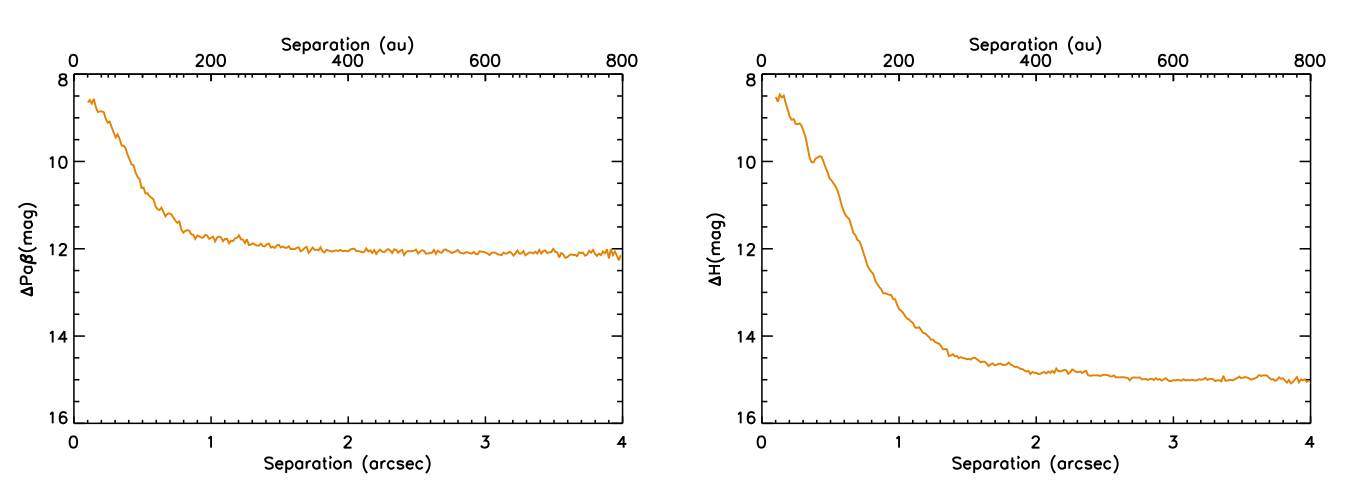}
\caption{{\it Left panel:} Contrast plot in magnitude obtained from the SHARK-NIR Pa$\beta$ data. {\it Right panel: } 
Contrast plot in magnitude obtained from the SHARK-NIR broad-band $H$ data.}
\label{f:contrastmag}
\end{figure*}

\begin{figure}
\centering
\includegraphics[width=0.49\textwidth]{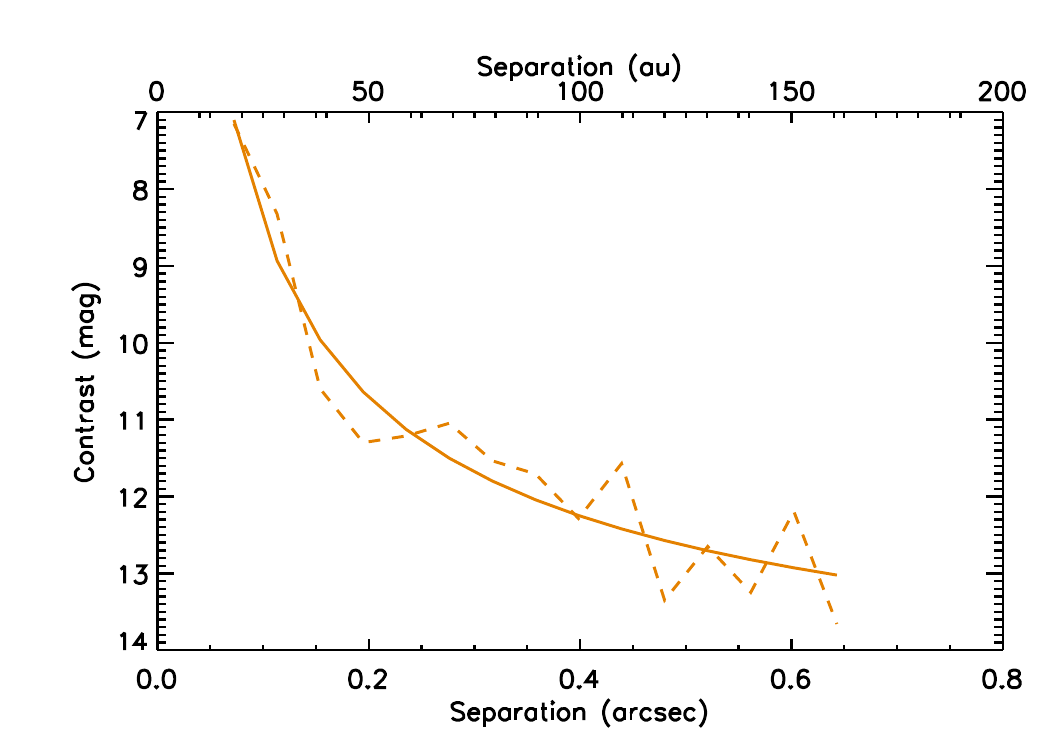}
\caption{Contrast plot obtained from the $H\alpha$ data of SHARK-VIS. The dashed line represents the contrast without any
smoothing while the solid line represents the smoothed version of the same contrast, useful to give the general behavior of
the contrast itself.}
\label{f:masslimitsVIS}
\end{figure}

\begin{figure}[!htp]
\centering
\includegraphics[width=0.49\textwidth]{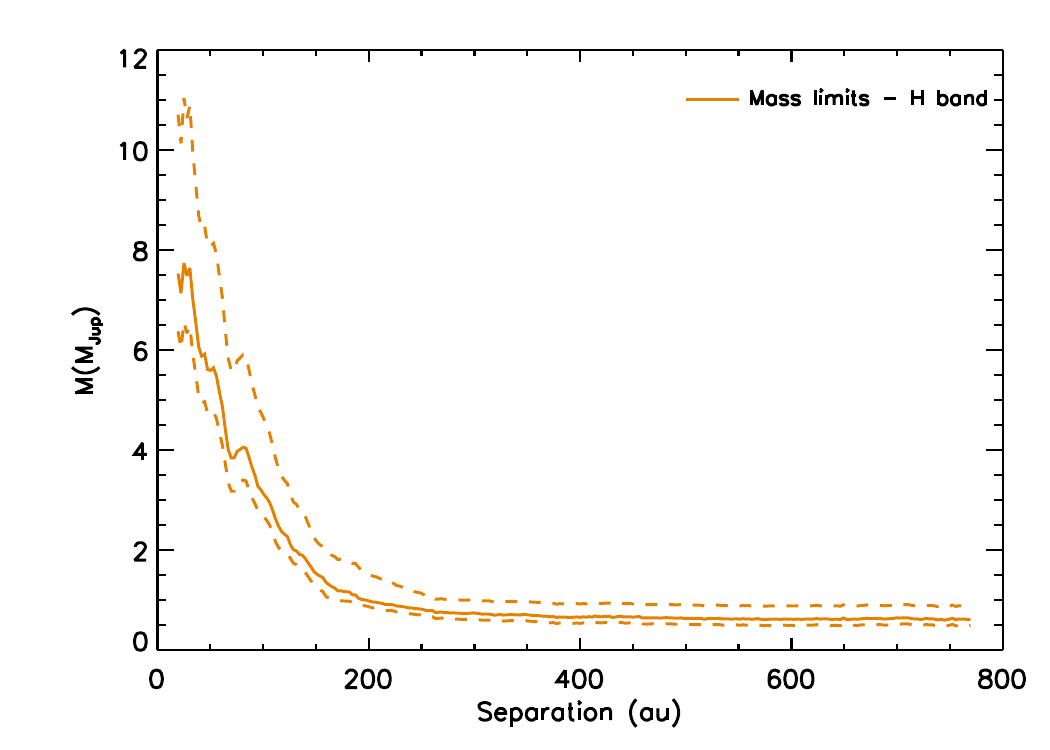}
\caption{Upper limits in mass expressed in \MJup obtained from the SHARK-NIR broad-band {\it H} data and adopting the
AMES-COND models. The two dashed lines denotes 
the mass limits obtained adopting the minimum (1~Myr) and maximum (3~Myr) ages for DR\,Tau. The solid line is obtained 
assuming an age 1.5~Myr.}
\label{f:masslimits}
\end{figure}

The $5\sigma$ contrast limits, as a function of the radial distance from the central star, were calculated for SHARK-NIR, calculating the standard deviation within one-pixel-wide annuli centered on the star. The self-subtraction caused by the post-processing method was evaluated by including simulated point-like sources of known brightness at different separations from the host star and corrected to obtain the final contrast. The point-like sources were generated as artificial PSF with a Gaussian shape and multiplied for the normalization factor obtained from the PSF outside the coronagraph described in Sec.~\ref{s:obs}. Finally, the results were corrected taking into account the effect of the small sample statistics as defined in \citet{2014ApJ...792...97M}. The final results from this procedure are displayed in the left panel of Fig.~\ref{f:contrastmag} for the {\it Pa$\beta$} image, while in the right panel we display the results from the broad-band~$H$ data. The contrast is much better in the latter case, because of both the better conditions of the second observing night (see Table~\ref{t:observation}) and the larger flux because of the larger bandpass. For the {\it Pa$\beta$} case, we obtain a contrast of 9.45~mag at a separation of 0\farcs3, of 11.75~mag at a separation of 1\as, and around 12~mag at separations above 1\as. However, the broad-band~$H$ data allow to reach a contrast of 9.31~mag at 0\farcs3, of 13.38~mag at 1\as, and greater than 14~mag at separations larger than 1\as. \par

For what concerns the SHARK-VIS observations, we evaluated the $5\sigma$ contrast limit in the {\it H$\alpha$} data using the 100-iterations PCA-IADI image. This was obtained by injecting simulated planets in the non-derotated frames. Our general procedure for computing contrast limits on the PCA-IADI result is based on \citet{2014ApJ...792...97M}. At first, we subdivided the PCA-IADI stack into a set of circular annuli and estimated the photometric noise as a function of the separation $r$ from the central star, $\sigma_P(r)$, by means of aperture photometry on a number of points angularly distributed along each annulus of the image. Then we injected a different number of simulated planets, generated with the same procedure described for SHARK-NIR, in each frame for each annulus. This number was chosen so that all the planets were at least two FWHM away from each other, and at least the full field angular range for the planets in same annulus. We computed the corresponding PCA-IADI with the same parameters used for the original frames. The simulated planets were injected with contrasts depending on r, namely proportional to the normalized radial profile $P(r)$ of the frames average stack, $C(\alpha, r)=\alpha\cdot P(r)$. We repeated this planet injection five times, ranging the proportionality coefficients $\alpha$ from 0.001 to 0.1 to have a set of contrast curves to interpolate. The contrast obtained from this procedure, smoothly interpolated to avoid the deviations due to the localized spiral arms in the PCA-IADI result and aimed to display the general behavior of the SHARK-VIS contrast, is displayed in Fig.~\ref{f:masslimitsVIS} with a solid orange line. In the same Figure we also display with a dashed orange line the non-smoothed version of the same contrast.

Assuming for the system the age and the distance given in Sec.~\ref{s:target} and using the AMES-COND models 
\citep{2003IAUS..211..325A} we then estimated the mass limits at different separations from the star. These results are displayed in Fig.~\ref{f:masslimits} for the broad-band~$H$ data. We do not display the same result for the {\it Pa$\beta$} because at that wavelength there could be some contribution from accreting companions. As it is not possible to disentangle between these two possible contributions, we decided to not use these data for the mass limits determination. 
For the broad-band~$H$ case, we obtain a mass limit of $\sim$7.64~\MJup at a separation of $\sim$30~au and of 3.11~\MJup at a separation of $\sim$100~au. \par
The same procedure applied to the visible data gives, as expected, much higher mass limits than for the NIR case. For this reason we do not display them in this work. 


\section{Discussion}
\label{s:dis}

\begin{figure}
\centering
\includegraphics[width=0.95\columnwidth]{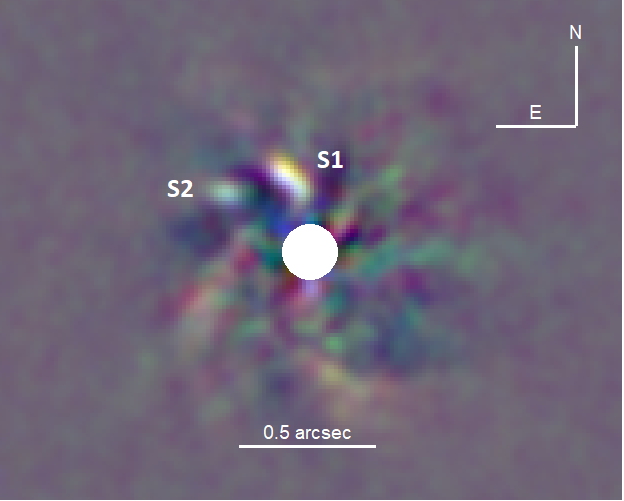}
\caption{Multi-color image obtained by overlapping the images from all the wavelength data. All the data are rescaled to the SHARK-NIR data. In the image, the blue is for the $H{\alpha}$ data, the green is for the Pa$\beta$ data, the red is for the broad-band~H data while the white is for flux in common for the three different wavelengths. The two structures S1 and S2 are also indicated. }
\label{f:colorimage}
\end{figure}

In Fig.~\ref{f:colorimage} we display a colored image obtained by overlaying the final frames obtained from all scientific filters used in this work. This image was obtained by using a blue color to show the SHARK-VIS data acquired with the {\it H${\alpha}$} filter, a green color for the SHARK-NIR data acquired using the narrow-band {\it Pa$\beta$} filter, and a red color for the SHARK-NIR data acquired using the broad-band~$H$ filter. All images are normalized to the peak of the PSF. A white color then implies similar flux ratios between a feature and the star for the three different wavelengths, which is roughly expected for reflected star light. The two compact structures described in Sec.~\ref{s:compstruct} and tagged with S1 and S2 appear whitish and a similar color is also obtained for other structures of the disk, namely the two spirals and the arc-like structure. This means that they have similar contrast in all the wavelengths considered. This suggests that their emission is mainly due to stellar light reflected by clumps of dust. Despite the suggestive detection of flux at the S1 and S2 positions using both the {\it H${\alpha}$} and the {\it Pa$\beta$} scientific filters, this probably does not come from accretion on compact objects or at least this is not true for the largest part of such emission. 

The structure tagged as S1 is clearly the same one described by \citet{2022A&A...658A..63M}, since both its separation and its PA are consistent with that described in that work. The S2 structure was not explicitly discussed in \citet{2022A&A...658A..63M} because its S/N was below the conventional threshold of 5, but it is also clearly visible in the SPHERE data. We notice that while the separation from the star of these structures is not changing, we have some indication of the variation of the PA. This will be discussed in the next subsection.

These two features were clearly recovered using three different instruments and with a time separation of around seven 
years. This should exclude the possibility that they are artifacts generated by the PCA methods used for the PSF
subtraction as it is very improbable that the same artifacts are created with different instruments and at such large 
time separation. It would be possible that they are created by self-subtraction of part of the disk structure as demonstrated for example by \citet{2017AJ....153..264F} and \citet{2018MNRAS.473.1774L}. However, like
for the case of the PSF subtraction artifacts, the probability that the same self-subtraction is generated in such different conditions is very low. We can then conclude that these features are real structures embedded in the 
DR\,Tau disk. A further possibility is that such structures could appear compact just at some wavelengths and not at other ones 
as highlighted for HD\,169142 in \citet{2018MNRAS.473.1774L}. However, we note that in our data the two structures appear very 
similar in all the filters used. Still, we cannot exclude that some difference in their structures could be present at 
wavelengths not used in our work.

Even in the case that they are due to reflected stellar light, they might be related to a planet formation process, similarly to what was proposed for the blob detected around HD\,169142 by \citet{2019A&A...623A.140G}. 

Finally, we notice that S1, with a projected separation of 58.7~au, is just at the border of the inner cavity of the dust disk observed by ALMA (program 2022.1.01365.S, PI Jane Huang). Such data were taken on August, 24 2022 in band 6 with an angular resolution of 0\farcs22 and with a continuum 
sensitivity of 0.0157 MJy\,beam$^{-1}$. The description of the data reduction of such data and the derivation of the main parameters of the DR\,Tau disk is given in \citet{2026A&A...715A..21G}. Indeed, in these data the disk size is $\sim$277~mas, corresponding to 53.5~au. On the contrary, S2, with a projected separation of 74.1~au, is well outside the disk border.

\subsection{Rotation of S1 and S2}

Both S1 and S2 features appear to rotate with time. The rate of rotation provides information about the source of the perturbation that causes them. Both S1 and S2 are likely extended sources, S1 being larger and more elongated than S2. The PA given for the SPHERE data in \citet{2022A&A...658A..63M} and those for the SHARK data given in Sec.~\ref{s:compstruct} have large error bars and it is difficult to calculate the rotation of the structures using them. We then determined the rotation rate by the peak of the cross-correlation of the images (in polar coordinates) obtained at different epochs in regions just around S1 and S2. Because of the low inclination of the disk that is very near to be face-on we did not deproject these images on the disk plane before to perform such cross-correlation.

We expect a difference in the position of the scattering surface at {\it H$\alpha$} with respect to the $H$-band due to the different penetration of photons of different wavelength. Given this concern, we evaluate the rotation of such structures only considering observations taken at the same wavelength ($H$-band) to estimate the rotation rates. From this procedure we find 
for S1 a total rotation of $4.58^{\circ}\pm0.67^{\circ}$ clockwise with respect to the IRDIS polarized data taken on November, 26 2018 for a total time range of 6.94~years. With respect to the same data, S2 has a rotation of $4.92^{\circ}\pm0.86^{\circ}$ in the same direction.

The rotation rates that we obtain this way are $0.66\pm 0.10$ degrees\,yr$^{-1}$ for S1 and $0.71\pm 0.12$ degrees\,yr$^{-1}$ 
for S2. The two values are in good agreement with each other. They are lower than indicated by the difference of 1.31 degrees between the PA of the polarimetric and SHINE SPHERE observations of S1 (acquired at a distance of about a year) obtained by \citet{2022A&A...658A..63M}. This may be due to the fact that in that case the rotation was calculated using SPHERE datasets taken at different wavelengths. The present results exploit a much longer baseline in time and are obtained at similar wavelengths. We then consider them to be more accurate. 

The agreement of the rotation rates obtained for S1 and S2 suggests that rotation of both sources is caused by the same perturbation. On the other hand, given the mass of $1.18^{+0.59}_{-0.44}$~\MSun \citep{2021ApJ...908...46B} of the star, the Keplerian rotation rates for circular orbits expected at the locations of S1 and S2 can be estimated applying the Kepler third law to calculate the period P(yr)$^2$=a(au)$^3$/M(\MSun) and dividing the round angle ($360^{\circ}$) for the period. Using this procedure we obtain $0.87\pm 0.19$ and $0.61\pm0.13$ degrees\,yr$^{-1}$, respectively. The rotation rates of S1 and S2 are then consistent with the Keplerian value at the location of S2. Although the observed rotation rates are also nominally consistent with that expected for the location of S1, they require a low value of 0.6~\MSun or less for the mass of the star, which is unlikely (see Sec.~\ref{s:target}). We conclude that the source of the perturbation is likely at the position of S2.

\subsection{Discussion about the possible presence of a planet within S2}
\label{s:possibleplanet}

\begin{figure*}
\centering
\includegraphics[width=0.95\textwidth]{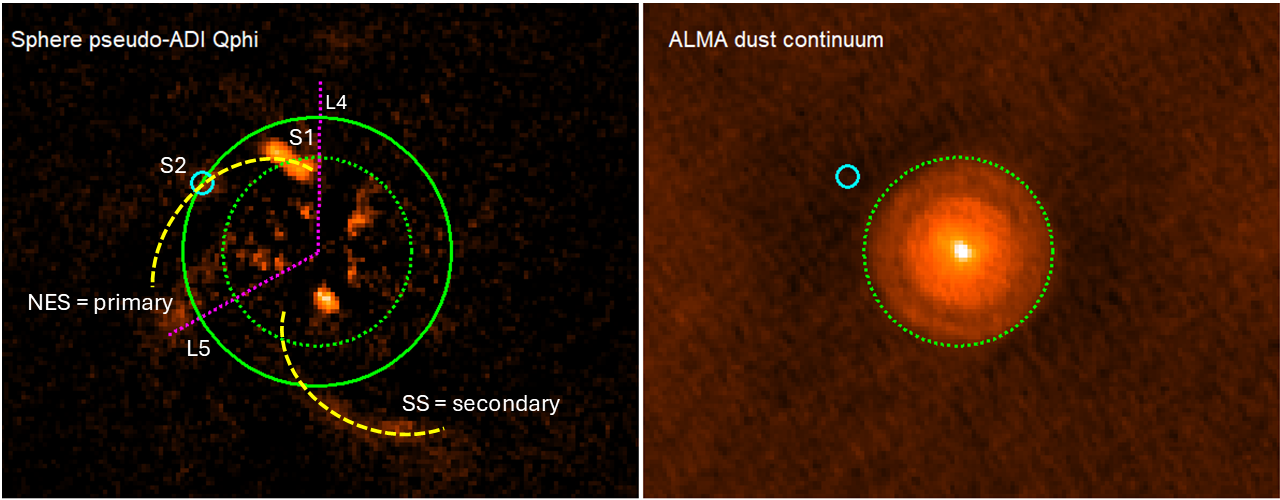}
\caption{{\it Left panel}: Pseudo-ADI polarimetric $Q_\phi$ image of DR Tau obtained from the SPHERE H broad band polarimetric data acquired in 2018. The position of S2 (at $\sim$74 au from the star) is marked with a cyan circle whose size approximately reproduces where the area with micron-sized dust is observed and a circular orbit with a solid green circle. The inner edge of the cavity observed in the ALMA continuum image is shown as a dashed green circle.  We also show vectors from the star rotated by $\pm$60 degrees with respect to S2; the intersections of these vectors with the orbit of S2 identify the L4 (leading) and L5 (trailing) Lagrangian points, where we might expect evidence of material co-orbiting S2. This might be expected in correspondence of the main phase of accretion on a planet located in S2. Finally, we signed the primary (NES) and the secondary spiral (SS) arm related to the disk perturbation if due to a planet at the location of S2. {\it Right panel:} the ALMA continuum image from band 6 data acquired within program 2022.1.01365.S (PI Jane Huang) on the same scale as for the SPHERE image. The expected position of S2 (cyan circle) takes into account the possible orbital motion around the star.}
\label{f:annotated}
\end{figure*}

In this section, we will show that several features of the DR Tau system fit in a scenario where S2 is a cloud that surrounds a planet in formation. To discuss this, we refer to the left panel of Fig.~\ref{f:annotated}, which shows a 
$Q_\phi$ image obtained from the SPHERE polarimetric data acquired in 2018 and presented in \citet{2022A&A...658A..63M}. The position of S2 (at $\sim$74 au from the star) is marked with a circle whose size approximately reproduces the area where micrometer dust is observed. We also marked the inner and outer edges of the cavity shown by the ALMA continuum image, whose inner and outer radii are 53.5 and 86 au, respectively. Given the faintness of the signal at large separation, the outer edge of the cavity is obtained from a median profile of the signal at a given separation. The magenta dotted lines are vectors from the star rotated by $\pm$60$^{\circ}$ with respect to S2; the intersections of these vectors with the orbit of S2 identify the L4 (leading) and L5 (trailing) Lagrangian points. Finally, the primary (NES) and the secondary spiral arm (SS) related to the disk perturbation are shown as yellow lines. The positions of such spiral arms are directly obtained from the position
of the maximum flux identified on the polarized image presented by \citet{2022A&A...658A..63M}.

These features explain most of what is observed in this image and can be interpreted in a semi-quantitative way. The location of S1 on the outer edge of the inner part of the disk seen by ALMA allows us to estimate precisely its distance from the star and then to estimate a pitch angle for the primary (NES) spiral arm of $18^{\circ}$ when we have (r/r$_p$)=0.7, where r is the distance from the star and r$_p$ is the distance between the star and S2. This is in good agreement with the value of $11^{\circ}$ estimated for the pitch angle of this spiral arm in \citet{2022A&A...658A..63M}. In agreement with what found for the DR\,Tau disk in \citet{2022A&A...658A..63M} we assume a linear perturbation regime, and using the results shown in Fig.~6 of \citet{Zhu2015} we can, given the pitch angle estimated above, calculate for the aspect ratio a value of (h/r)=0.15, where h is the height of the disk at a distance r from the star. Using this value of (h/r) and assuming a value of 0.001 for the viscosity parameter, a gap with a width of W=32.5~au, as determined using the ALMA data, created by a planet located at r$_p$=74.1~au is expected from an object with a mass ratio of q=0.0021 \citep{Kanagawa2016}.  
In addition, the secondary (SS) spiral arm is at an angle of 105.5$^{\circ}$, as seen by the star from the Primary (NES) one; this implies $q$=0.0012 using the relation by \citet{Fung2015}. In consideration of the rough approximation done, these different estimates agree with each other, suggesting that the proposed planet might have a mass of about 1-2~\MJup.

\begin{figure}
\centering
\includegraphics[width=0.95\columnwidth]{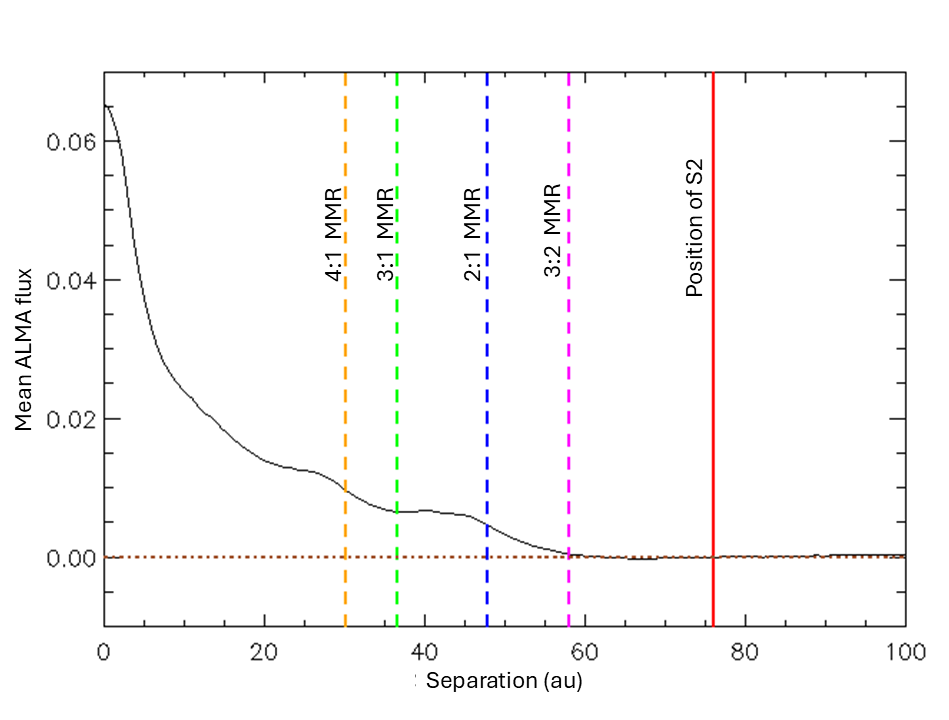}
\caption{Radial profile of the ALMA dust continuum image. The dashed vertical lines with different colours are at the expected separations for the 3:2, 2:1, 3:1, and 4:1 MMR with the possible planet at the S2 location that is indicated by the solid red line.}
\label{f:ALMA_profile}
\end{figure}

\begin{figure*}[!htp]
\centering
\sidecaption
\includegraphics[width=12cm]{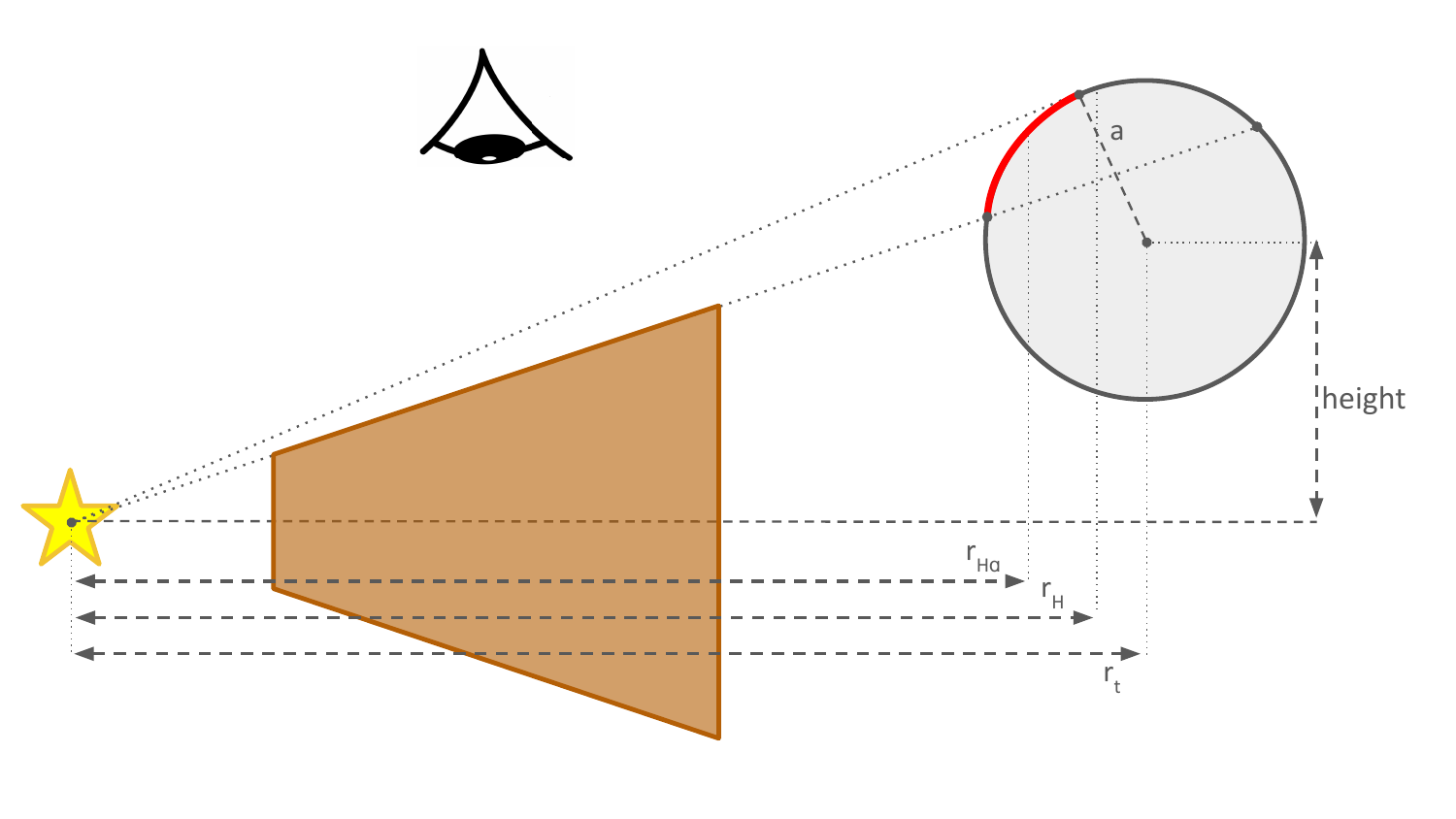}
\caption{Diagram representation of the visibility of the S2 structure considering the disk shadow and its height with respect to the disk plane.
Observer, indicated by the eye, is from the top of the model. The dimensions are not in scale.}
\label{f:disksketch}
\end{figure*}

We also notice that mean motion resonances (MMR) with a planet located at S2 may well justify the presence of rings and gaps in the ALMA dust image, as shown by a comparison with the radial profile given in Fig.~\ref{f:ALMA_profile}. The ALMA dust disk is all contained within the 3:2 MMR that matches its outer edge. Peaks of the profile (rings) are seen close to the 2:1 and 4:1 MMR, while the 3:1 MMR corresponds to a minimum (gap). Planet clearing gaps are often invoked as a possible explanation for gaps in disks; however, until now the results do not provide strong evidence to support or reject this conclusion \citep{vanderMarel2019}. If it is possible to confirm the presence of a planetary mass companion in S2,
DR\,Tau might then be among the first evidences for this.

These arguments suggest the possibility that an object with a mass of about 1-2 \MJup is related to S2. In principle, we might assume that light from S2 originates from the photosphere of a planet. From the contrasts given in Sec.~\ref{s:compstruct} we could derive for such object an absolute magnitudes of 12.18 for the broad-band~$H$ data. These
magnitudes may be reproduced by a planet with a mass of 2.2 or 3.3 \MJup if we adopt an age of 1 Myr and the COND 
\citep{Baraffe1998} and DUSTY \citep{Chabrier2000} models, respectively. 

Given the uncertainties, this is compatible with the values of the mass determined from dynamical methods. However, this interpretation is largely incompatible with the observation in {\it H$\alpha$} for two reasons. First, the expected $R$ magnitude of a 1 Myr old 2 \MJup planet is much fainter than that observed. However, we are not considering here the possible flux coming for the accretion on a proto-planet. Considering it properly could  possibly overcome this problem. But the second main issue is that the position of S2 is slightly but significantly different when observed in {\it H$\alpha$} (separation of 362$\pm$8~mas) and in the broad $H$ band (separation of 384$\pm$7~mas). This difference cannot be explained if S2 is a point source, but it may be understood as an effect of optical depth (the opacity being larger at shorter wavelengths) if S2 is an extended cloud reflecting stellar light. Of course, this requires that the radius of the cloud is larger than the offset between the position of S2 in the {\it H$\alpha$} and $H$ band, that is, at least 22$\pm$7 mas. 

\begin{table*}[]
\caption{Real position and size of S2 as a function of the height with respect to the disk plane given in the first column. }
\centering
\begin{tabular}{cccccccc}
\hline \hline
Height	&\multicolumn{2}{c}{$r_t$}			&	\multicolumn{2}{c}{$a$}			&	Size {\it H} band			&	Size {\it H$\alpha$} band& $f$ \\		
(au)	&	(mas)	&	(au)	&	(mas)	&	(au)	&		(mas)			&	(mas)	\\	
\hline
0	&	393.5	&	75.9	&	74.0	&	14.3	&		88.8	$\times$	90.3	&	41.9	$\times$	90.3	&0.00046\\
5	&	390	&	75.3	&	54.5	&	10.5	&		86.1	$\times$	87.6	&	39.2	$\times$	85.4	&0.00070\\
10	&	387	&	74.7	&	39.5	&	7.6	&		76.9	$\times$	77.7	&	32.1	$\times$	76.6	&0.00104\\
15	&	384	&	74.1	&	32.0	&	6.2	&		63.8	$\times$	63.8	&	26.0	$\times$	61.2	&0.00151\\
\hline
\end{tabular}
\label{t:s2}
\end{table*}

We notice that to our knowledge such difference in separation has not been measured previously for other objects. This is not completely surprising
given that at the moment such structures detected embedded in protoplanetary disks are relatively rare and this effect should also
depend from the geometry of the system being favored in disk seen almost face-on. For this reason,
with the aim to explain in a more quantitative way the offset between the position of S2 in {\it H$\alpha$} and in broad-band~$H$ data, we developed a toy model that we illustrate in the following. This model can work independently from the fact if S2 is simply a dust cloud in the DR\,Tau disk or if it is containing a forming planet. To this aim, we consider a model of S2 as an extended cloud that is farther from the star than the inner part of the disk revealed by ALMA, so that only a part of it is likely illuminated by the star because of the shadow casted by the inner part of the disk. This may lead to offsets in its apparent position that depend on the exact geometry and on the optical depth of the cloud. The situation is visually described in Fig.~\ref{f:disksketch}.  The dimensions in the Figure are not in scale and its aim is simply to show how the model is working. In this Figure $r_t$ represents the real distance projected on the disk plane of the S2 structure that is assumed to be spherical and with a radius a. The disk (indicated in red in the Figure) shadow allows to see just a little part of the structure. Assuming that S2 is thick with respect to $H\alpha$ radiation we are able to see just the part of the cloud marked in red. At such wavelength we then see S2 at a distance from the star defined r$_{H\alpha}$. On the contrary, assuming the cloud thin with respect to $H$ radiation, we can see the part marked with red and grey line and, as a consequence, we see the structure at a different distance from the star defined r$_H$. \citet{2007A&A...469..963P} developed a model with stratified layers for a disk that can further validate the
proposed model. In this model, the visible radiation would be scattered at the surface of the disk where small grains reside. On the
contrary, NIR radiation could reach deeper layers of the disk where larger grains resides. Following this model, the S2 cloud would
be composed in its external layers of small grains that could scatter visible wavelengths justifying the assumption of the optical 
thickness at such wavelengths. Larger grains would instead be stratified in the inner part of the cloud. 
This toy model is able to explain the different separation from the star that we measure when observing at different wavelengths. A similar difference in optical depth at different wavelengths could be considered also for the inner disk leading to a different shadow pattern according to the wavelength. However, we should remind the very large dimension of the disk itself
extending up to 58.7~au. This makes very improbable that radiation of any wavelength can effectively pass through its extension and for this reason in our model we treated it as a ''solid'' object casting the same shadow at all the wavelengths. \par
To quantify the offset, we considered some rough approximations. We assumed that the disk is opaque with a scale height (h/r)=0.15 (see previous discussion) independent of wavelength, that S2 is a sphere with a radius a located at a true distance r$_t$ from the star as projected on the disk plane, and that it is severely optically thick at $H\alpha$ and optically thin in the $H$ band. We also neglected the inclination of the disk with respect to the line of sight, so that we could use the observed quantities directly. Given the low inclination of the DR\,Tau disk, this is a good approximation. We considered different heights of the center of S2 with respect to the disk plane, in the range from 0 to 15 au; for the last value, S2 is fully out of the disk shadow. We considered solutions that match the observed values of the apparent separations in H$\alpha$ (362 mas) and in the H band (384 mas). The real position and size of S2 as a function of its height above the disk plane are given in Table \ref{t:s2}. Here, $r_t$ represents the real separation, given both in angular separation and in au, of the center of the S2 structure while a is the radius of the same structure considered circular. The column 6 and 7 are giving the apparent extension in mas for the S2 structure when observing with the broad-band~$H$ and $H\alpha$ filter, respectively.
Finally, in the last column $f$ represents the fraction of stellar light intercepted by the S2 structure.We also give the total apparent size of the illuminated regions of S2 along both the radial and the tangential directions in Columns 6 and 7 when using the broad-band~$H$ and the narrow-band $H\alpha$ filters, respectively. Within these approximations, the image of S2 is smaller in $H\alpha$ because we assume that only (part of) the surface illuminated by the star is visible to us, while photons in the $H$ band penetrate inward and also illuminate the far side of the cloud. The last column of Table \ref{t:s2} gives the fraction of starlight that is intercepted by S2. This may be compared with the observed contrast. If, for example, we adopt an albedo of 0.1, the expected contrast in the $H\alpha$ image if the center of S2 is in the disk plane is $4.6\times 10^{-5}$, which coincides with the value given in Sec.~\ref{s:compstruct}. This suggests that the visible/near infrared image of S2 can be explained with an extended cloud that is optically thin in the $H$-band and optically thick in $H\alpha$.

We also notice that there is no detectable source in the position of S2 in the ALMA band 6 dust continuum data (see the right panel of Fig.~\ref{f:annotated}). A 2$\sigma$ upper limit to the flux in this position (computed using the variance of the flux of similar areas at the same distance from the star) is $1.78\times 10^{-3}$ mJy at 240 GHz (ALMA band 6).  
On the other hand, a comparison with Fig.~4 of \citet{Facchini2026} indicates that the upper limit to ALMA continuum flux for S2 is compatible with the flux expected from a circumplanetary disk around a planet with a mass $<1.9$ \MJup. This means that we do not really expect to detect in the ALMA image a circumplanetary disk around a planet located in S2 that is possibly responsible of the dynamical perturbation. 

Although we cannot definitely show that this is the only explanation of all observations, the scenario that considers a giant planet in its formation phase at the location of S2 then reproduces most of the observations. Traditionally, core accretion is not expected to form a giant planet around a solar type star as young as DR\,Tau (age $\sim$1~Myr) and at a separation as large as 74~au \citep{Rafikov2011, Helled2014}. However, scenarios where giant planets form very early at large separations by pebble accretion have been proposed recently. For example, \citet{Gurrutxaga2024} found that, under suitable conditions, planet cores can grow up to their pebble isolation mass, between approximately 20 and 80~M$_\oplus$, within less than 1 Myr, in the 20-80~au region around solar-like stars. Such planets are expected to later have a fast inward migration in the early stages of gas accretion, towards the $\sim10$ au region. The occurrence of these wide-orbit gas giants should be relatively low because their core must attain sufficient mass to accrete gas before the pebble flux decays, while avoiding excessive migration. In the case of S2, we should observe the planet, if it exists, just during the early phase of the gas accretion, after the end of pebble accretion (as indicated by the lack of emission from the millimetric grains at the planet separation) and before migration moved it much closer to the star. \par
Alternatively, the planet, if it exists, might have formed through gravitational instability \citep{1998ApJ...503..923B,2013A&A...552A.129V} that allows a quick formation of planetary mass objects in the outer part
of the disks from the contraction of clumps detaching from the rest of the disk.

{\subsection{Possible alternative explanations}
\label{s:alt}

In the previous Sections we have discussed the possible presence of a forming planet embedded in the DR\,Tau disk showing that 
it could explain our data but that we cannot demonstrate its existence without any doubt. Anyway, our data could also be 
explained through alternative mechanisms. In this Section we will discuss them briefly. Such a discussion was partially
developed for DR\,Tau in \citet{2022A&A...658A..63M}. We will further develop such discussion in the following. \par
One of the possible causes for the spiral arms in a disk is the gravitational instability. We can take as reference for 
this mechanism the work by \citet{2015ApJ...812L..32D}. In Table~1 of that paper they calculated the expected rotation for 
different values of $q=M_{disk}/M_*$. For the lowest case that they considered ($q$=0.125) they found a rotation of $\sim$0.9$^{\circ}$/year, a  
value not so different from the rotation value found from our data. However, they also found that the gravitational instability 
require that the spiral pattern is quite compact with an extension less than 100~au while the spiral arms in DR\,Tau extend out to 
220~au from the star. Moreover, they also concluded that the gravitational instability needs a value of $q$ larger than 0.1 while 
\citet{2022A&A...658A..63M} evaluated
a value of $q$$\sim$0.03 well below that limit. Updated parameters in our work, especially the stellar mass, could slightly
change that value but not in a sensitive way and probably toward still lower values. \citet{2022A&A...658A..63M} also used an 
estimation of the value of the Toomre parameter \citep[Q;][]{1964ApJ...139.1217T} to establish if the disk could be instable. 
Indeed, a value of Q larger than 1.7 is a clear indication of a stable disk. They calculated for Q a value of 3, well into the 
stability range. They used for that calculation a value of 0.1 for the aspect ratio while in this work we estimated a value of 
0.15 so that the estimated of the value of Q could change but again not towards a disk in which the gravitational instability is 
acting. All these considerations can then help us to exclude that the gravitational instability can be the cause of what we see in 
the DR\,Tau disk. \par
A second mechanism that could generate a spiral pattern is the passage of a perturbing body at a separation less than 
1000~au in a recent past. This possibility was also explored in \citet{2022A&A...658A..63M} that found that the star 
DQ\,Tau have passed at a separation of $\sim$5000~au from DR\,Tau about 0.231$\pm$0.007~Myr ago while other passages at 
shorter separations but much more time ago were also individuated. Anyway, as highlighted in \citet{2014MNRAS.441.2094R}, this 
material should have dispersed on a large spatial range not allowing to see anymore the spiral pattern 
after a time range of around 0.1~Myr. For this reason we can consider very improbable that this mechanism is the 
responsible for the DR\,Tau spiral pattern. \par 
A further mechanism that was explored in \citet{2022A&A...658A..63M} was the late infall of material on the DR\,Tau disk 
following the models proposed by \citet{2019A&A...628A..20D} and \citet{2020A&A...633A...3K}. They actually found that this
is a mechanism that can perfectly explain the formation of a spiral model and that the interstellar 
environment around DR\,Tau could supply the material needed to sustain this mechanism. We also notice that the simulation 
of \citet{2020A&A...633A...3K} and \citet{2025MNRAS.537.2695C} support keplerian rotation of the disk at least for some particular initial conditions. Also, 
the models do not foresee that all the material is going in the disk pattern but some of this could be lost and appear as an
arc-like structures. This structure is very similar to that tagged as ALS in \citet{2022A&A...658A..63M} and also recovered in 
our data. Finally, recent models developed by \citet{2025MNRAS.537.2695C} for late infall of material led by accretion 
streamers allowed to obtain results comparable to the spiral pattern imaged by \citet{2022A&A...658A..63M}. This mechanism is 
then a very reliable alternative explanations for the spiral arm formation and rotation for DR\,Tau. 


\section{Conclusions}
\label{s:conclusion}

In this work, we presented observations of the young star DR\,Tau taken in parallel in the visible and the NIR wavelengths using the LBT high-contrast imagers SHARK-VIS and SHARK-NIR. In our images, we were able to retrieve the disk structures, namely two spiral arms and one arc-like structure, previously discovered around this star, even if at very low S/N. This is expected as we are not observing this target using polarized light, which was fundamental in previous SPHERE observations 
\citep{2022A&A...658A..63M} to image the disk at high S/N. In any case, our observations were able to detect and confirm all the main structures previously detected for this disk.

The main objective of our observation was to detect possible point-like sources embedded in the disk. In particular, the observations in $H{\alpha}$ with SHARK-VIS and in $Pa\beta$ with SHARK-NIR were aimed at detecting possible objects experiencing accretion. We were able to detect two structures tagged S1 and S2 in both the narrow-band scientific filters and also in broad-band~$H$. The structure tagged S1 appears to be elongated while S2 is a more compact structure. The S1 structure can be identified with one structure found in the SPHERE data taken almost seven years before our observations \citep{2022A&A...658A..63M}. Its separation and position angle are clearly consistent with those found with SPHERE. However, we found a possible clockwise rotation of $\sim$4.5$^{\circ}$. On the other hand, the S2
structure was not considered in the previous analysis using SPHERE data because its S/N was below the threshold of 5. In any case, it was clearly identifiable also in those data and we could then calculate its rotation finding a value of $\sim$4.9$^{\circ}$ in the same direction than S1.   
Despite the presence from both the structures described above of flux at $H{\alpha}$ and $Pa\beta$ wavelengths, we are probably not observing flux caused by accretion. Instead it is likely stellar light reflected from dust indicating the presence of large dust clumps as demonstrated by the similar contrast in all the wavelength bands considered in this work. \par
The rotation rate calculated for both these structures is very similar to each other and compatible with a keplerian rotation for an object at the separation of S2. These results give hints for a rotation led by an object at the separation of S2. This might possibly be caused by the presence of an embedded planet caught during its formation process with a possible mass of a few \MJup. However, this is just a suggestive hypothesis that our data cannot confirm without any doubt. Alternative mechanisms have been explored and they could explain the spiral pattern of the
DR\,Tau disk even if the presence of the two embedded structures is harder to explain.\par

These results confirm the very interesting nature of DR\,Tau, and give some hints about the possible formation of planets in progress in the disk around this star. If this will be confirmed, it will be the first time that it will
be observed the planetary formation process at such an early stage. While very suggestive, these data do not
allow to draw definitive conclusions about the nature of such objects. Further observations and analysis 
at different wavelengths and with different instruments will be needed to further confirm or disproof such hypothesis.


\begin{acknowledgements}
The authors thank the anonymous referee for the fundamental help in improving this work. We thank Tom Herbst from MPIA-Heidelberg and the LINC-NIRVANA team for 
sharing part of their instrument control software to operate the motorized axis of 
SHARK-NIR. We also express our appreciation to NASA and Marcia Rieke, the Principal Investigator of JWST/NIRCam, for granting us the opportunity to utilize one of the NIRCam spare detectors as the primary detector for the SHARK-NIR scientific camera. Additionally, we acknowledge the significant contribution of the ALTA Center (alta.arcetri.inaf.it) forecasts, generated using the Astro-Meso-Nh model. Initialization data of the ALTA automatic forecast system is sourced from the
General Circulation Model (HRES) of the European Centre for Medium-Range Weather Forecasts.
We gratefully acknowledge support from the “Programma di Ricerca Fondamentale INAF 2023” of the Italian National Institute of Astrophysics (INAF Large Grant 2023 “NextSTEPS”).
The LBT is an international collaboration among institutions in the United States, Italy and Germany. The LBT Corporation partners are: The University of Arizona on behalf of the Arizona university system; Istituto Nazionale di Astrofisica, Italy; LBT Beteiligungsgesellschaft, Germany, representing the Max Planck Society, the Astrophysical Institute Potsdam, and Heidelberg University; The Ohio State University; The Research Corporation, on behalf of The University of Notre Dame, University of Minnesota and University of Virginia.
This work has made use of data from the European Space Agency (ESA) mission
{\it Gaia} (\url{https://www.cosmos.esa.int/gaia}), processed by
the {\it Gaia} Data Processing and Analysis Consortium (DPAC,
\url{https://www.cosmos.esa.int/web/gaia/dpac/consortium}). Funding for
the DPAC has been provided by national institutions, in particular, the
institutions participating in the {\it Gaia} Multilateral Agreement.
This research has made extensive use of the NASA-ADS, SIMBAD and Vizier database, operated at CDS, Strasbourg, France.
VD and SA acknowledges the financial contribution from PRIN-MUR 2022YP5ACE. AZ acknowledges support from ANID -- Millennium Science Initiative Program -- Center Code NCN2024\_001 and Fondecyt Regular grant number 1250249.
This paper makes use of the following ALMA data: ADS/JAO.ALMA 2022.1.01365.S ALMA is a partnership of ESO (representing its member states), NSF (USA) and NINS (Japan), together with NRC (Canada), NSTC and ASIAA (Taiwan), and KASI (Republic of Korea), in cooperation with the Republic of Chile. The Joint ALMA Observatory is operated by ESO, AUI/NRAO and NAOJ.
\end{acknowledgements}

\bibliographystyle{aa}
\bibliography{drtau}

@ARTICLE{2023ApJ...952L..17W,
       author = {{Weber}, Philipp and {P{\'e}rez}, Sebasti{\'a}n and {Zurlo}, Alice and {Miley}, James and {Hales}, Antonio and {Cieza}, Lucas and {Principe}, David and {C{\'a}rcamo}, Miguel and {Garufi}, Antonio and {K{\'o}sp{\'a}l}, {\'A}gnes and {Takami}, Michihiro and {Kastner}, Joel and {Zhu}, Zhaohuan and {Williams}, Jonathan},
        title = "{Spirals and Clumps in V960 Mon: Signs of Planet Formation via Gravitational Instability around an FU Ori Star?}",
      journal = {\apjl},
         year = 2023,
        month = jul,
       volume = {952},
       number = {1},
          eid = {L17},
        pages = {L17},
          doi = {10.3847/2041-8213/ace186},
archivePrefix = {arXiv},
       eprint = {2307.13433},
 primaryClass = {astro-ph.EP},
       adsurl = {https://ui.adsabs.harvard.edu/abs/2023ApJ...952L..17W}
}

@ARTICLE{2025ApJ...988L..30D,
       author = {{Dasgupta}, Anuroop and {Zurlo}, Alice and {Weber}, Philipp and {Maio}, Francesco and {Cieza}, Lucas A. and {Fedele}, Davide and {Garufi}, Antonio and {Miley}, James and {Pathak}, Prashant and {P{\'e}rez}, Sebasti{\'a}n and {Roccatagliata}, Veronica},
        title = "{VLT/ERIS Observations of the V960 Mon System: A Dust-embedded Substellar Object Formed by Gravitational Instability?}",
      journal = {\apjl},
         year = 2025,
        month = jul,
       volume = {988},
       number = {1},
          eid = {L30},
        pages = {L30},
          doi = {10.3847/2041-8213/ade996},
archivePrefix = {arXiv},
       eprint = {2507.13571},
 primaryClass = {astro-ph.SR},
       adsurl = {https://ui.adsabs.harvard.edu/abs/2025ApJ...988L..30D}
}

@ARTICLE{2021A&A...649A...1G,
       author = {{Gaia Collaboration} and {Brown}, A.~G.~A. and {Vallenari}, A. and {Prusti}, T. and {de Bruijne}, J.~H.~J. and {Babusiaux}, C. and {Biermann}, M. and {Creevey}, O.~L. and {Evans}, D.~W. and {Eyer}, L. and {Hutton}, A. and {Jansen}, F. and {Jordi}, C. and {Klioner}, S.~A. and {Lammers}, U. and {Lindegren}, L. and {Luri}, X. and {Mignard}, F. and {Panem}, C. and {Pourbaix}, D. and {Randich}, S. and {Sartoretti}, P. and {Soubiran}, C. and {Walton}, N.~A. and {Arenou}, F. and {Bailer-Jones}, C.~A.~L. and {Bastian}, U. and {Cropper}, M. and {Drimmel}, R. and {Katz}, D. and {Lattanzi}, M.~G. and {van Leeuwen}, F. and {Bakker}, J. and {Cacciari}, C. and {Casta{\~n}eda}, J. and {De Angeli}, F. and {Ducourant}, C. and {Fabricius}, C. and {Fouesneau}, M. and {Fr{\'e}mat}, Y. and {Guerra}, R. and {Guerrier}, A. and {Guiraud}, J. and {Jean-Antoine Piccolo}, A. and {Masana}, E. and {Messineo}, R. and {Mowlavi}, N. and {Nicolas}, C. and {Nienartowicz}, K. and {Pailler}, F. and {Panuzzo}, P. and {Riclet}, F. and {Roux}, W. and {Seabroke}, G.~M. and {Sordo}, R. and {Tanga}, P. and {Th{\'e}venin}, F. and {Gracia-Abril}, G. and {Portell}, J. and {Teyssier}, D. and {Altmann}, M. and {Andrae}, R. and {Bellas-Velidis}, I. and {Benson}, K. and {Berthier}, J. and {Blomme}, R. and {Brugaletta}, E. and {Burgess}, P.~W. and {Busso}, G. and {Carry}, B. and {Cellino}, A. and {Cheek}, N. and {Clementini}, G. and {Damerdji}, Y. and {Davidson}, M. and {Delchambre}, L. and {Dell'Oro}, A. and {Fern{\'a}ndez-Hern{\'a}ndez}, J. and {Galluccio}, L. and {Garc{\'\i}a-Lario}, P. and {Garcia-Reinaldos}, M. and {Gonz{\'a}lez-N{\'u}{\~n}ez}, J. and {Gosset}, E. and {Haigron}, R. and {Halbwachs}, J.-L. and {Hambly}, N.~C. and {Harrison}, D.~L. and {Hatzidimitriou}, D. and {Heiter}, U. and {Hern{\'a}ndez}, J. and {Hestroffer}, D. and {Hodgkin}, S.~T. and {Holl}, B. and {Jan{\ss}en}, K. and {Jevardat de Fombelle}, G. and {Jordan}, S. and {Krone-Martins}, A. and {Lanzafame}, A.~C. and {L{\"o}ffler}, W. and {Lorca}, A. and {Manteiga}, M. and {Marchal}, O. and {Marrese}, P.~M. and {Moitinho}, A. and {Mora}, A. and {Muinonen}, K. and {Osborne}, P. and {Pancino}, E. and {Pauwels}, T. and {Petit}, J.-M. and {Recio-Blanco}, A. and {Richards}, P.~J. and {Riello}, M. and {Rimoldini}, L. and {Robin}, A.~C. and {Roegiers}, T. and {Rybizki}, J. and {Sarro}, L.~M. and {Siopis}, C. and {Smith}, M. and {Sozzetti}, A. and {Ulla}, A. and {Utrilla}, E. and {van Leeuwen}, M. and {van Reeven}, W. and {Abbas}, U. and {Abreu Aramburu}, A. and {Accart}, S. and {Aerts}, C. and {Aguado}, J.~J. and {Ajaj}, M. and {Altavilla}, G. and {{\'A}lvarez}, M.~A. and {{\'A}lvarez Cid-Fuentes}, J. and {Alves}, J. and {Anderson}, R.~I. and {Anglada Varela}, E. and {Antoja}, T. and {Audard}, M. and {Baines}, D. and {Baker}, S.~G. and {Balaguer-N{\'u}{\~n}ez}, L. and {Balbinot}, E. and {Balog}, Z. and {Barache}, C. and {Barbato}, D. and {Barros}, M. and {Barstow}, M.~A. and {Bartolom{\'e}}, S. and {Bassilana}, J.-L. and {Bauchet}, N. and {Baudesson-Stella}, A. and {Becciani}, U. and {Bellazzini}, M. and {Bernet}, M. and {Bertone}, S. and {Bianchi}, L. and {Blanco-Cuaresma}, S. and {Boch}, T. and {Bombrun}, A. and {Bossini}, D. and {Bouquillon}, S. and {Bragaglia}, A. and {Bramante}, L. and {Breedt}, E. and {Bressan}, A. and {Brouillet}, N. and {Bucciarelli}, B. and {Burlacu}, A. and {Busonero}, D. and {Butkevich}, A.~G. and {Buzzi}, R. and {Caffau}, E. and {Cancelliere}, R. and {C{\'a}novas}, H. and {Cantat-Gaudin}, T. and {Carballo}, R. and {Carlucci}, T. and {Carnerero}, M.~I. and {Carrasco}, J.~M. and {Casamiquela}, L. and {Castellani}, M. and {Castro-Ginard}, A. and {Castro Sampol}, P. and {Chaoul}, L. and {Charlot}, P. and {Chemin}, L. and {Chiavassa}, A. and {Cioni}, M.-R.~L. and {Comoretto}, G. and {Cooper}, W.~J. and {Cornez}, T. and {Cowell}, S. and {Crifo}, F. and {Crosta}, M. and {Crowley}, C. and {Dafonte}, C. and {Dapergolas}, A. and {David}, M. and {David}, P.},
        title = "{Gaia Early Data Release 3. Summary of the contents and survey properties}",
      journal = {\aap},
         year = 2021,
        month = may,
       volume = {649},
          eid = {A1},
        pages = {A1},
          doi = {10.1051/0004-6361/202039657},
archivePrefix = {arXiv},
       eprint = {2012.01533},
 primaryClass = {astro-ph.GA},
       adsurl = {https://ui.adsabs.harvard.edu/abs/2021A&A...649A...1G}
}

@ARTICLE{2001AJ....122.3335A,
       author = {{Alencar}, Silvia H.~P. and {Johns-Krull}, Christopher M. and {Basri}, Gibor},
        title = "{The Spectral Variability of the Classical T Tauri Star DR Tauri}",
      journal = {\aj},
         year = 2001,
        month = dec,
       volume = {122},
       number = {6},
        pages = {3335-3360},
          doi = {10.1086/323914},
       adsurl = {https://ui.adsabs.harvard.edu/abs/2001AJ....122.3335A}
}

@ARTICLE{2007A&A...461..183G,
       author = {{Grankin}, K.~N. and {Melnikov}, S. Yu. and {Bouvier}, J. and {Herbst}, W. and {Shevchenko}, V.~S.},
        title = "{Results of the ROTOR-program. I. The long-term photometric variability of classical T Tauri stars}",
      journal = {\aap},
         year = 2007,
        month = jan,
       volume = {461},
       number = {1},
        pages = {183-195},
          doi = {10.1051/0004-6361:20065489},
archivePrefix = {arXiv},
       eprint = {astro-ph/0611028},
 primaryClass = {astro-ph},
       adsurl = {https://ui.adsabs.harvard.edu/abs/2007A&A...461..183G}
}

@ARTICLE{2019A&A...632A..32M,
       author = {{McClure}, M.~K.},
        title = "{Carbon depletion observed inside T Tauri inner rims. Formation of icy, kilometer size planetesimals by 1 Myr}",
      journal = {\aap},
         year = 2019,
        month = dec,
       volume = {632},
          eid = {A32},
        pages = {A32},
          doi = {10.1051/0004-6361/201834361},
archivePrefix = {arXiv},
       eprint = {1910.06029},
 primaryClass = {astro-ph.SR},
       adsurl = {https://ui.adsabs.harvard.edu/abs/2019A&A...632A..32M}
}

@ARTICLE{2011A&A...535A...6P,
       author = {{Petrov}, P.~P. and {Gahm}, G.~F. and {Stempels}, H.~C. and {Walter}, F.~M. and {Artemenko}, S.~A.},
        title = "{Accretion-powered chromospheres in classical T Tauri stars}",
      journal = {\aap},
         year = 2011,
        month = nov,
       volume = {535},
          eid = {A6},
        pages = {A6},
          doi = {10.1051/0004-6361/201116721},
archivePrefix = {arXiv},
       eprint = {1109.1266},
 primaryClass = {astro-ph.SR},
       adsurl = {https://ui.adsabs.harvard.edu/abs/2011A&A...535A...6P}
}

@ARTICLE{2019ApJ...882...49L,
       author = {{Long}, Feng and {Herczeg}, Gregory J. and {Harsono}, Daniel and {Pinilla}, Paola and {Tazzari}, Marco and {Manara}, Carlo F. and {Pascucci}, Ilaria and {Cabrit}, Sylvie and {Nisini}, Brunella and {Johnstone}, Doug and {Edwards}, Suzan and {Salyk}, Colette and {Menard}, Francois and {Lodato}, Giuseppe and {Boehler}, Yann and {Mace}, Gregory N. and {Liu}, Yao and {Mulders}, Gijs D. and {Hendler}, Nathanial and {Ragusa}, Enrico and {Fischer}, William J. and {Banzatti}, Andrea and {Rigliaco}, Elisabetta and {van de Plas}, Gerrit and {Dipierro}, Giovanni and {Gully-Santiago}, Michael and {Lopez-Valdivia}, Ricardo},
        title = "{Compact Disks in a High-resolution ALMA Survey of Dust Structures in the Taurus Molecular Cloud}",
      journal = {\apj},
         year = 2019,
        month = sep,
       volume = {882},
       number = {1},
          eid = {49},
        pages = {49},
          doi = {10.3847/1538-4357/ab2d2d},
archivePrefix = {arXiv},
       eprint = {1906.10809},
 primaryClass = {astro-ph.SR},
       adsurl = {https://ui.adsabs.harvard.edu/abs/2019ApJ...882...49L}
}

@ARTICLE{2022A&A...667A.124G,
       author = {{Gangi}, M. and {Antoniucci}, S. and {Biazzo}, K. and {Frasca}, A. and {Nisini}, B. and {Alcal{\'a}}, J.~M. and {Giannini}, T. and {Manara}, C.~F. and {Giunta}, A. and {Harutyunyan}, A. and {Munari}, U. and {Vitali}, F.},
        title = "{GIARPS High-resolution Observations of T Tauri stars (GHOsT). IV. Accretion properties of the Taurus-Auriga young association}",
      journal = {\aap},
         year = 2022,
        month = nov,
       volume = {667},
          eid = {A124},
        pages = {A124},
          doi = {10.1051/0004-6361/202244042},
archivePrefix = {arXiv},
       eprint = {2208.14895},
 primaryClass = {astro-ph.SR},
       adsurl = {https://ui.adsabs.harvard.edu/abs/2022A&A...667A.124G}
}

@ARTICLE{2021ApJ...908...46B,
       author = {{Braun}, Teresa A.~M. and {Yen}, Hsi-Wei and {Koch}, Patrick M. and {Manara}, Carlo F. and {Miotello}, Anna and {Testi}, Leonardo},
        title = "{Dynamical Stellar Masses of Pre-main-sequence Stars in Lupus and Taurus Obtained with ALMA Surveys in Comparison with Stellar Evolutionary Models}",
      journal = {\apj},
         year = 2021,
        month = feb,
       volume = {908},
       number = {1},
          eid = {46},
        pages = {46},
          doi = {10.3847/1538-4357/abd24f},
archivePrefix = {arXiv},
       eprint = {2012.07441},
 primaryClass = {astro-ph.SR},
       adsurl = {https://ui.adsabs.harvard.edu/abs/2021ApJ...908...46B}
}

@ARTICLE{2013ApJ...770L..14B,
       author = {{Brown}, Logan R. and {Troutman}, Matthew R. and {Gibb}, Erika L.},
        title = "{Spectro-astrometry of Water in DR Tauri}",
      journal = {\apjl},
         year = 2013,
        month = jun,
       volume = {770},
       number = {1},
          eid = {L14},
        pages = {L14},
          doi = {10.1088/2041-8205/770/1/L14},
       adsurl = {https://ui.adsabs.harvard.edu/abs/2013ApJ...770L..14B}
}

@ARTICLE{2024A&A...686A.117T,
       author = {{Temmink}, Milou and {van Dishoeck}, Ewine F. and {Grant}, Sierra L. and {Tabone}, Beno{\^\i}t and {Gasman}, Danny and {Christiaens}, Valentin and {Samland}, Matthias and {Argyriou}, Ioannis and {Perotti}, Giulia and {G{\"u}del}, Manuel and {Henning}, Thomas and {Lagage}, Pierre-Olivier and {Abergel}, Alain and {Absil}, Olivier and {Barrado}, David and {Caratti o Garatti}, Alessio and {Glauser}, Adrian M. and {Kamp}, Inga and {Lahuis}, Fred and {Olofsson}, G{\"o}ran and {Ray}, Tom P. and {Scheithauer}, Silvia and {Vandenbussche}, Bart and {Waters}, L.~B.~F.~M. and {Arabhavi}, Aditya M. and {Jang}, Hyerin and {Kanwar}, Jayatee and {Morales-Calder{\'o}n}, Maria and {Rodgers-Lee}, Donna and {Schreiber}, J{\"u}rgen and {Schwarz}, Kamber and {Colina}, Luis},
        title = "{MINDS: The DR Tau disk. I. Combining JWST-MIRI data with high-resolution CO spectra to characterise the hot gas}",
      journal = {\aap},
         year = 2024,
        month = jun,
       volume = {686},
          eid = {A117},
        pages = {A117},
          doi = {10.1051/0004-6361/202348911},
archivePrefix = {arXiv},
       eprint = {2403.13591},
 primaryClass = {astro-ph.EP},
       adsurl = {https://ui.adsabs.harvard.edu/abs/2024A&A...686A.117T}
}

@ARTICLE{2024A&A...689A.330T,
       author = {{Temmink}, Milou and {van Dishoeck}, Ewine F. and {Gasman}, Danny and {Grant}, Sierra L. and {Tabone}, Beno{\^\i}t and {G{\"u}del}, Manuel and {Henning}, Thomas and {Barrado}, David and {Caratti o Garatti}, Alessio and {Glauser}, Adrian M. and {Kamp}, Inga and {Arabhavi}, Aditya M. and {Jang}, Hyerin and {Kurtovic}, Nicolas and {Perotti}, Giulia and {Schwarz}, Kamber and {Vlasblom}, Marissa},
        title = "{MINDS: The DR Tau disk: II. Probing the hot and cold H$_{2}$O reservoirs in the JWST-MIRI spectrum}",
      journal = {\aap},
         year = 2024,
        month = sep,
       volume = {689},
          eid = {A330},
        pages = {A330},
          doi = {10.1051/0004-6361/202450355},
archivePrefix = {arXiv},
       eprint = {2407.05070},
 primaryClass = {astro-ph.EP},
       adsurl = {https://ui.adsabs.harvard.edu/abs/2024A&A...689A.330T}
}

@ARTICLE{2023ApJ...943..107H,
       author = {{Huang}, Jane and {Bergin}, Edwin A. and {Bae}, Jaehan and {Benisty}, Myriam and {Andrews}, Sean M.},
        title = "{Molecular Mapping of DR Tau's Protoplanetary Disk, Envelope, Outflow, and Large-scale Spiral Arm}",
      journal = {\apj},
         year = 2023,
        month = feb,
       volume = {943},
       number = {2},
          eid = {107},
        pages = {107},
          doi = {10.3847/1538-4357/aca89c},
archivePrefix = {arXiv},
       eprint = {2301.02674},
 primaryClass = {astro-ph.SR},
       adsurl = {https://ui.adsabs.harvard.edu/abs/2023ApJ...943..107H}
}

@ARTICLE{2022A&A...658A..63M,
       author = {{Mesa}, D. and {Ginski}, C. and {Gratton}, R. and {Ertel}, S. and {Wagner}, K. and {Bonavita}, M. and {Fedele}, D. and {Meyer}, M. and {Henning}, T. and {Langlois}, M. and {Garufi}, A. and {Antoniucci}, S. and {Claudi}, R. and {Defr{\`e}re}, D. and {Desidera}, S. and {Janson}, M. and {Pawellek}, N. and {Rigliaco}, E. and {Squicciarini}, V. and {Zurlo}, A. and {Boccaletti}, A. and {Bonnefoy}, M. and {Cantalloube}, F. and {Chauvin}, G. and {Feldt}, M. and {Hagelberg}, J. and {Hugot}, E. and {Lagrange}, A.-M. and {Lazzoni}, C. and {Maurel}, D. and {Perrot}, C. and {Petit}, C. and {Rouan}, D. and {Vigan}, A.},
        title = "{Signs of late infall and possible planet formation around DR Tau using VLT/SPHERE and LBTI/LMIRCam}",
      journal = {\aap},
         year = 2022,
        month = feb,
       volume = {658},
          eid = {A63},
        pages = {A63},
          doi = {10.1051/0004-6361/202142219},
archivePrefix = {arXiv},
       eprint = {2111.01702},
 primaryClass = {astro-ph.EP},
       adsurl = {https://ui.adsabs.harvard.edu/abs/2022A&A...658A..63M}
}

@ARTICLE{2015IJAsB..14..365F,
       author = {{Farinato}, Jacopo and {Baffa}, Carlo and {Baruffolo}, Andrea and {Bergomi}, Maria and {Carbonaro}, Luca and {Carlotti}, Alexis and {Centrone}, Mauro and {Codona}, Johanan and {Dima}, Marco and {Esposito}, Simone and {Fantinel}, Daniela and {Farisato}, Giancarlo and {Gaessler}, Wolfgang and {Giallongo}, Emanuele and {Greggio}, Davide and {Hinz}, Philip and {Lisi}, Franco and {Magrin}, Demetrio and {Marafatto}, Luca and {Pedichini}, Fernando and {Pinna}, Enrico and {Puglisi}, Alfio and {Ragazzoni}, Roberto and {Salasnich}, Bernardo and {Stangalini}, Marco and {Verinaud}, Christophe and {Viotto}, Valentina},
        title = "{The NIR arm of SHARK: System for coronagraphy with High-order Adaptive optics from R to K bands}",
      journal = {International Journal of Astrobiology},
         year = 2015,
        month = jul,
       volume = {14},
       number = {3},
        pages = {365-373},
          doi = {10.1017/S1473550414000494},
       adsurl = {https://ui.adsabs.harvard.edu/abs/2015IJAsB..14..365F}
}

@INPROCEEDINGS{2022SPIE12185E..22F,
       author = {{Farinato}, Jacopo and {Baruffolo}, Andrea and {Bergomi}, Maria and {Bianco}, Andrea and {Biondi}, Federico and {Briegel}, Florian and {Carolo}, Elena and {Carlotti}, Alexis and {Chavan}, Sona and {Chinellato}, Simonetta and {De Pascale}, Marco and {Dima}, Marco and {D'Orazi}, Valentina and {Ertel}, Steve and {Greggio}, Davide and {Henning}, Thomas and {Laudisio}, Fulvio and {Lessio}, Luigi and {Magrin}, Demetrio and {Marafatto}, Luca and {Mesa}, Dino and {Mohr}, Lars and {Montoya}, Manny and {Radhakrishnan}, Kalyan and {Ricci}, Davide and {Umbriaco}, Gabriele and {Vassallo}, Daniele and {Viotto}, Valentina and {Zanutta}, Alessio and {Antoniucci}, Simone and {Arcidiacono}, Carmelo and {Bacciotti}, Francesca and {Baudoz}, Pierre and {Bongiorno}, Angela and {Close}, Laird and {Di Filippo}, Simone and {Don}, Kenneth and {Esposito}, Simone and {Grenz}, Paul and {Guyon}, Olivier and {Leisenring}, Jarron M. and {Pedichini}, Fernando and {Piazzesi}, Roberto and {Pinna}, Enrico and {Portaluri}, Elisa and {Puglisi}, Alfio and {Ragazzoni}, Roberto and {Rossi}, Fabio},
        title = "{SHARK-NIR, ready to ``swim'' in the LBT Northern Hemisphere ``ocean''}",
    booktitle = {Adaptive Optics Systems VIII},
         year = 2022,
       editor = {{Schreiber}, Laura and {Schmidt}, Dirk and {Vernet}, Elise},
       series = {Society of Photo-Optical Instrumentation Engineers (SPIE) Conference Series},
       volume = {12185},
        month = aug,
          eid = {1218522},
        pages = {1218522},
          doi = {10.1117/12.2630083},
       adsurl = {https://ui.adsabs.harvard.edu/abs/2022SPIE12185E..22F}
}

@INPROCEEDINGS{2022SPIE12184E..3VM,
       author = {{Marafatto}, L. and {Carolo}, E. and {Umbriaco}, G. and {Bergomi}, M. and {Greggio}, D. and {Laudisio}, F. and {Lessio}, L. and {Montoya}, M. and {Radhakrishnan}, K. and {Ricci}, D. and {Vassallo}, D. and {Dima}, M. and {Di Filippo}, S. and {Don}, K.~W.~W. and {Viotto}, V. and {Farinato}, J.},
        title = "{SHARK-NIR on its way to LBT}",
    booktitle = {Ground-based and Airborne Instrumentation for Astronomy IX},
         year = 2022,
       editor = {{Evans}, Christopher J. and {Bryant}, Julia J. and {Motohara}, Kentaro},
       series = {Society of Photo-Optical Instrumentation Engineers (SPIE) Conference Series},
       volume = {12184},
        month = aug,
          eid = {121843V},
        pages = {121843V},
          doi = {10.1117/12.2629494},
       adsurl = {https://ui.adsabs.harvard.edu/abs/2022SPIE12184E..3VM}
}

@ARTICLE{2006ApJ...641..556M,
       author = {{Marois}, Christian and {Lafreni{\`e}re}, David and {Doyon}, Ren{\'e} and {Macintosh}, Bruce and {Nadeau}, Daniel},
        title = "{Angular Differential Imaging: A Powerful High-Contrast Imaging Technique}",
      journal = {\apj},
         year = 2006,
        month = apr,
       volume = {641},
       number = {1},
        pages = {556-564},
          doi = {10.1086/500401},
archivePrefix = {arXiv},
       eprint = {astro-ph/0512335},
 primaryClass = {astro-ph},
       adsurl = {https://ui.adsabs.harvard.edu/abs/2006ApJ...641..556M}
}

@ARTICLE{2025MNRAS.536.1455M,
       author = {{Mesa}, D. and {Gratton}, R. and {D'Orazi}, V. and {Carolo}, E. and {Vassallo}, D. and {Farinato}, J. and {Marafatto}, L. and {Wagner}, K. and {Hom}, J. and {Ertel}, S. and {Henning}, Th and {Desgrange}, C. and {Barbato}, D. and {Bergomi}, M. and {Cerpelloni}, P. and {Desidera}, S. and {Di Filippo}, S. and {Doelman}, D. and {Machado}, T.~S. Gomes and {Greggio}, D. and {Grenz}, P. and {Kenworthy}, M. and {Laudisio}, F. and {Lazzoni}, C. and {Leisenring}, J. and {Lorenzetto}, A. and {Santhakumari}, K.~K.~R. and {Ricci}, D. and {Snik}, F. and {Umbriaco}, G. and {Pallauta}, M.~C. Vega and {Viotto}, V.},
        title = "{Deep imaging of three accelerating stars using SHARK-NIR and LMIRCam at LBT}",
      journal = {\mnras},
         year = 2025,
        month = jan,
       volume = {536},
       number = {2},
        pages = {1455-1466},
          doi = {10.1093/mnras/stae2706},
archivePrefix = {arXiv},
       eprint = {2412.04171},
 primaryClass = {astro-ph.EP},
       adsurl = {https://ui.adsabs.harvard.edu/abs/2025MNRAS.536.1455M}
}

@ARTICLE{2012MNRAS.427..948A,
       author = {{Amara}, Adam and {Quanz}, Sascha P.},
        title = "{PYNPOINT: an image processing package for finding exoplanets}",
      journal = {\mnras},
         year = 2012,
        month = dec,
       volume = {427},
       number = {2},
        pages = {948-955},
          doi = {10.1111/j.1365-2966.2012.21918.x},
archivePrefix = {arXiv},
       eprint = {1207.6637},
 primaryClass = {astro-ph.IM},
       adsurl = {https://ui.adsabs.harvard.edu/abs/2012MNRAS.427..948A}
}

@ARTICLE{2012ApJ...755L..28S,
       author = {{Soummer}, R{\'e}mi and {Pueyo}, Laurent and {Larkin}, James},
        title = "{Detection and Characterization of Exoplanets and Disks Using Projections on Karhunen-Lo{\`e}ve Eigenimages}",
      journal = {\apjl},
         year = 2012,
        month = aug,
       volume = {755},
       number = {2},
          eid = {L28},
        pages = {L28},
          doi = {10.1088/2041-8205/755/2/L28},
archivePrefix = {arXiv},
       eprint = {1207.4197},
 primaryClass = {astro-ph.IM},
       adsurl = {https://ui.adsabs.harvard.edu/abs/2012ApJ...755L..28S}
}

@ARTICLE{2011A&A...528L..15B,
       author = {{Bonnefoy}, M. and {Lagrange}, A.-M. and {Boccaletti}, A. and {Chauvin}, G. and {Apai}, D. and {Allard}, F. and {Ehrenreich}, D. and {Girard}, J.~H.~V. and {Mouillet}, D. and {Rouan}, D. and {Gratadour}, D. and {Kasper}, M.},
        title = "{High angular resolution detection of {\ensuremath{\beta}} Pictoris b at 2.18 {\ensuremath{\mu}}m}",
      journal = {\aap},
         year = 2011,
        month = apr,
       volume = {528},
          eid = {L15},
        pages = {L15},
          doi = {10.1051/0004-6361/201016224},
       adsurl = {https://ui.adsabs.harvard.edu/abs/2011A&A...528L..15B}
}

@ARTICLE{2014A&A...572A..85Z,
       author = {{Zurlo}, A. and {Vigan}, A. and {Mesa}, D. and {Gratton}, R. and {Moutou}, C. and {Langlois}, M. and {Claudi}, R.~U. and {Pueyo}, L. and {Boccaletti}, A. and {Baruffolo}, A. and {Beuzit}, J.-L. and {Costille}, A. and {Desidera}, S. and {Dohlen}, K. and {Feldt}, M. and {Fusco}, T. and {Henning}, T. and {Kasper}, M. and {Martinez}, P. and {Moeller-Nilsson}, O. and {Mouillet}, D. and {Pavlov}, A. and {Puget}, P. and {Sauvage}, J.-F. and {Turatto}, M. and {Udry}, S. and {Vakili}, F. and {Waters}, R. and {Wildi}, R.~F.},
        title = "{Performance of the VLT Planet Finder SPHERE. I. Photometry and astrometry precision with IRDIS and IFS in laboratory}",
      journal = {\aap},
         year = 2014,
        month = dec,
       volume = {572},
          eid = {A85},
        pages = {A85},
          doi = {10.1051/0004-6361/201424204},
archivePrefix = {arXiv},
       eprint = {1410.1754},
 primaryClass = {astro-ph.EP},
       adsurl = {https://ui.adsabs.harvard.edu/abs/2014A&A...572A..85Z}
}

@ARTICLE{2014ApJ...792...97M,
       author = {{Mawet}, D. and {Milli}, J. and {Wahhaj}, Z. and {Pelat}, D. and {Absil}, O. and {Delacroix}, C. and {Boccaletti}, A. and {Kasper}, M. and {Kenworthy}, M. and {Marois}, C. and {Mennesson}, B. and {Pueyo}, L.},
        title = "{Fundamental Limitations of High Contrast Imaging Set by Small Sample Statistics}",
      journal = {\apj},
         year = 2014,
        month = sep,
       volume = {792},
       number = {2},
          eid = {97},
        pages = {97},
          doi = {10.1088/0004-637X/792/2/97},
archivePrefix = {arXiv},
       eprint = {1407.2247},
 primaryClass = {astro-ph.IM},
       adsurl = {https://ui.adsabs.harvard.edu/abs/2014ApJ...792...97M}
}

@INPROCEEDINGS{2003IAUS..211..325A,
       author = {{Allard}, France and {Guillot}, Tristan and {Ludwig}, Hans-G{\"u}nter and {Hauschildt}, Peter H. and {Schweitzer}, Andreas and {Alexander}, David R. and {Ferguson}, Jason W.},
        title = "{Model Atmospheres and Spectra: The Role of Dust}",
    booktitle = {Brown Dwarfs},
         year = 2003,
       editor = {{Mart{\'\i}n}, Eduardo},
       series = {IAU Symposium},
       volume = {211},
        month = jun,
        pages = {325},
       adsurl = {https://ui.adsabs.harvard.edu/abs/2003IAUS..211..325A}
}

@ARTICLE{2012ApJ...756..133C,
       author = {{Chen}, Christine H. and {Pecaut}, Mark and {Mamajek}, Eric E. and {Su}, Kate Y.~L. and {Bitner}, Martin},
        title = "{A Spitzer MIPS Study of 2.5-2.0 M $_{{\ensuremath{\odot}}}$ Stars in Scorpius-Centaurus}",
      journal = {\apj},
         year = 2012,
        month = sep,
       volume = {756},
       number = {2},
          eid = {133},
        pages = {133},
          doi = {10.1088/0004-637X/756/2/133},
archivePrefix = {arXiv},
       eprint = {1207.3415},
 primaryClass = {astro-ph.SR},
       adsurl = {https://ui.adsabs.harvard.edu/abs/2012ApJ...756..133C}
}

@ARTICLE{2014A&A...565A..15M,
       author = {{Marshall}, J.~P. and {Moro-Mart{\'\i}n}, A. and {Eiroa}, C. and {Kennedy}, G. and {Mora}, A. and {Sibthorpe}, B. and {Lestrade}, J.-F. and {Maldonado}, J. and {Sanz-Forcada}, J. and {Wyatt}, M.~C. and {Matthews}, B. and {Horner}, J. and {Montesinos}, B. and {Bryden}, G. and {del Burgo}, C. and {Greaves}, J.~S. and {Ivison}, R.~J. and {Meeus}, G. and {Olofsson}, G. and {Pilbratt}, G.~L. and {White}, G.~J.},
        title = "{Correlations between the stellar, planetary, and debris components of exoplanet systems observed by Herschel}",
      journal = {\aap},
         year = 2014,
        month = may,
       volume = {565},
          eid = {A15},
        pages = {A15},
          doi = {10.1051/0004-6361/201323058},
archivePrefix = {arXiv},
       eprint = {1403.6186},
 primaryClass = {astro-ph.EP},
       adsurl = {https://ui.adsabs.harvard.edu/abs/2014A&A...565A..15M}
}

@ARTICLE{2018A&A...617A..44K,
       author = {{Keppler}, M. and {Benisty}, M. and {M{\"u}ller}, A. and {Henning}, Th. and {van Boekel}, R. and {Cantalloube}, F. and {Ginski}, C. and {van Holstein}, R.~G. and {Maire}, A.-L. and {Pohl}, A. and {Samland}, M. and {Avenhaus}, H. and {Baudino}, J.-L. and {Boccaletti}, A. and {de Boer}, J. and {Bonnefoy}, M. and {Chauvin}, G. and {Desidera}, S. and {Langlois}, M. and {Lazzoni}, C. and {Marleau}, G.-D. and {Mordasini}, C. and {Pawellek}, N. and {Stolker}, T. and {Vigan}, A. and {Zurlo}, A. and {Birnstiel}, T. and {Brandner}, W. and {Feldt}, M. and {Flock}, M. and {Girard}, J. and {Gratton}, R. and {Hagelberg}, J. and {Isella}, A. and {Janson}, M. and {Juhasz}, A. and {Kemmer}, J. and {Kral}, Q. and {Lagrange}, A.-M. and {Launhardt}, R. and {Matter}, A. and {M{\'e}nard}, F. and {Milli}, J. and {Molli{\`e}re}, P. and {Olofsson}, J. and {P{\'e}rez}, L. and {Pinilla}, P. and {Pinte}, C. and {Quanz}, S.~P. and {Schmidt}, T. and {Udry}, S. and {Wahhaj}, Z. and {Williams}, J.~P. and {Buenzli}, E. and {Cudel}, M. and {Dominik}, C. and {Galicher}, R. and {Kasper}, M. and {Lannier}, J. and {Mesa}, D. and {Mouillet}, D. and {Peretti}, S. and {Perrot}, C. and {Salter}, G. and {Sissa}, E. and {Wildi}, F. and {Abe}, L. and {Antichi}, J. and {Augereau}, J.-C. and {Baruffolo}, A. and {Baudoz}, P. and {Bazzon}, A. and {Beuzit}, J.-L. and {Blanchard}, P. and {Brems}, S.~S. and {Buey}, T. and {De Caprio}, V. and {Carbillet}, M. and {Carle}, M. and {Cascone}, E. and {Cheetham}, A. and {Claudi}, R. and {Costille}, A. and {Delboulb{\'e}}, A. and {Dohlen}, K. and {Fantinel}, D. and {Feautrier}, P. and {Fusco}, T. and {Giro}, E. and {Gluck}, L. and {Gry}, C. and {Hubin}, N. and {Hugot}, E. and {Jaquet}, M. and {Le Mignant}, D. and {Llored}, M. and {Madec}, F. and {Magnard}, Y. and {Martinez}, P. and {Maurel}, D. and {Meyer}, M. and {M{\"o}ller-Nilsson}, O. and {Moulin}, T. and {Mugnier}, L. and {Orign{\'e}}, A. and {Pavlov}, A. and {Perret}, D. and {Petit}, C. and {Pragt}, J. and {Puget}, P. and {Rabou}, P. and {Ramos}, J. and {Rigal}, F. and {Rochat}, S. and {Roelfsema}, R. and {Rousset}, G. and {Roux}, A. and {Salasnich}, B. and {Sauvage}, J.-F. and {Sevin}, A. and {Soenke}, C. and {Stadler}, E. and {Suarez}, M. and {Turatto}, M. and {Weber}, L.},
        title = "{Discovery of a planetary-mass companion within the gap of the transition disk around PDS 70}",
      journal = {\aap},
         year = 2018,
        month = sep,
       volume = {617},
          eid = {A44},
        pages = {A44},
          doi = {10.1051/0004-6361/201832957},
archivePrefix = {arXiv},
       eprint = {1806.11568},
 primaryClass = {astro-ph.EP},
       adsurl = {https://ui.adsabs.harvard.edu/abs/2018A&A...617A..44K}
}

@ARTICLE{2019NatAs...3..749H,
       author = {{Haffert}, S.~Y. and {Bohn}, A.~J. and {de Boer}, J. and {Snellen}, I.~A.~G. and {Brinchmann}, J. and {Girard}, J.~H. and {Keller}, C.~U. and {Bacon}, R.},
        title = "{Two accreting protoplanets around the young star PDS 70}",
      journal = {Nature Astronomy},
         year = 2019,
        month = jun,
       volume = {3},
        pages = {749-754},
          doi = {10.1038/s41550-019-0780-5},
archivePrefix = {arXiv},
       eprint = {1906.01486},
 primaryClass = {astro-ph.EP},
       adsurl = {https://ui.adsabs.harvard.edu/abs/2019NatAs...3..749H}
}

@ARTICLE{2017ApJ...850...52P,
       author = {{Pohl}, A. and {Benisty}, M. and {Pinilla}, P. and {Ginski}, C. and {de Boer}, J. and {Avenhaus}, H. and {Henning}, Th. and {Zurlo}, A. and {Boccaletti}, A. and {Augereau}, J.-C. and {Birnstiel}, T. and {Dominik}, C. and {Facchini}, S. and {Fedele}, D. and {Janson}, M. and {Keppler}, M. and {Kral}, Q. and {Langlois}, M. and {Ligi}, R. and {Maire}, A.-L. and {M{\'e}nard}, F. and {Meyer}, M. and {Pinte}, C. and {Quanz}, S.~P. and {Sauvage}, J.-F. and {Sezestre}, {\'E}. and {Stolker}, T. and {Szul{\'a}gyi}, J. and {van Boekel}, R. and {van der Plas}, G. and {Villenave}, M. and {Baruffolo}, A. and {Baudoz}, P. and {Le Mignant}, D. and {Maurel}, D. and {Ramos}, J. and {Weber}, L.},
        title = "{The Circumstellar Disk HD 169142: Gas, Dust, and Planets Acting in Concert?}",
      journal = {\apj},
         year = 2017,
        month = nov,
       volume = {850},
       number = {1},
          eid = {52},
        pages = {52},
          doi = {10.3847/1538-4357/aa94c2},
archivePrefix = {arXiv},
       eprint = {1710.06485},
 primaryClass = {astro-ph.EP},
       adsurl = {https://ui.adsabs.harvard.edu/abs/2017ApJ...850...52P}
}

@ARTICLE{2018MNRAS.473.1774L,
       author = {{Ligi}, R. and {Vigan}, A. and {Gratton}, R. and {de Boer}, J. and {Benisty}, M. and {Boccaletti}, A. and {Quanz}, S.~P. and {Meyer}, M. and {Ginski}, C. and {Sissa}, E. and {Gry}, C. and {Henning}, T. and {Beuzit}, J.-L. and {Biller}, B. and {Bonnefoy}, M. and {Chauvin}, G. and {Cheetham}, A.~C. and {Cudel}, M. and {Delorme}, P. and {Desidera}, S. and {Feldt}, M. and {Galicher}, R. and {Girard}, J. and {Janson}, M. and {Kasper}, M. and {Kopytova}, T. and {Lagrange}, A.-M. and {Langlois}, M. and {Lecoroller}, H. and {Maire}, A.-L. and {M{\'e}nard}, F. and {Mesa}, D. and {Peretti}, S. and {Perrot}, C. and {Pinilla}, P. and {Pohl}, A. and {Rouan}, D. and {Stolker}, T. and {Samland}, M. and {Wahhaj}, Z. and {Wildi}, F. and {Zurlo}, A. and {Buey}, T. and {Fantinel}, D. and {Fusco}, T. and {Jaquet}, M. and {Moulin}, T. and {Ramos}, J. and {Suarez}, M. and {Weber}, L.},
        title = "{Investigation of the inner structures around HD 169142 with VLT/SPHERE}",
      journal = {\mnras},
         year = 2018,
        month = jan,
       volume = {473},
       number = {2},
        pages = {1774-1783},
          doi = {10.1093/mnras/stx2318},
archivePrefix = {arXiv},
       eprint = {1709.01734},
 primaryClass = {astro-ph.EP},
       adsurl = {https://ui.adsabs.harvard.edu/abs/2018MNRAS.473.1774L}
}

@ARTICLE{2019A&A...623A.140G,
       author = {{Gratton}, R. and {Ligi}, R. and {Sissa}, E. and {Desidera}, S. and {Mesa}, D. and {Bonnefoy}, M. and {Chauvin}, G. and {Cheetham}, A. and {Feldt}, M. and {Lagrange}, A.~M. and {Langlois}, M. and {Meyer}, M. and {Vigan}, A. and {Boccaletti}, A. and {Janson}, M. and {Lazzoni}, C. and {Zurlo}, A. and {De Boer}, J. and {Henning}, T. and {D'Orazi}, V. and {Gluck}, L. and {Madec}, F. and {Jaquet}, M. and {Baudoz}, P. and {Fantinel}, D. and {Pavlov}, A. and {Wildi}, F.},
        title = "{Blobs, spiral arms, and a possible planet around HD 169142}",
      journal = {\aap},
         year = 2019,
        month = mar,
       volume = {623},
          eid = {A140},
        pages = {A140},
          doi = {10.1051/0004-6361/201834760},
archivePrefix = {arXiv},
       eprint = {1901.06555},
 primaryClass = {astro-ph.EP},
       adsurl = {https://ui.adsabs.harvard.edu/abs/2019A&A...623A.140G}
}

@ARTICLE{2022NatAs...6..751C,
       author = {{Currie}, Thayne and {Lawson}, Kellen and {Schneider}, Glenn and {Lyra}, Wladimir and {Wisniewski}, John and {Grady}, Carol and {Guyon}, Olivier and {Tamura}, Motohide and {Kotani}, Takayuki and {Kawahara}, Hajime and {Brandt}, Timothy and {Uyama}, Taichi and {Muto}, Takayuki and {Dong}, Ruobing and {Kudo}, Tomoyuki and {Hashimoto}, Jun and {Fukagawa}, Misato and {Wagner}, Kevin and {Lozi}, Julien and {Chilcote}, Jeffrey and {Tobin}, Taylor and {Groff}, Tyler and {Ward-Duong}, Kimberly and {Januszewski}, William and {Norris}, Barnaby and {Tuthill}, Peter and {van der Marel}, Nienke and {Sitko}, Michael and {Deo}, Vincent and {Vievard}, Sebastien and {Jovanovic}, Nemanja and {Martinache}, Frantz and {Skaf}, Nour},
        title = "{Images of embedded Jovian planet formation at a wide separation around AB Aurigae}",
      journal = {Nature Astronomy},
         year = 2022,
        month = apr,
       volume = {6},
        pages = {751-759},
          doi = {10.1038/s41550-022-01634-x},
archivePrefix = {arXiv},
       eprint = {2204.00633},
 primaryClass = {astro-ph.EP},
       adsurl = {https://ui.adsabs.harvard.edu/abs/2022NatAs...6..751C}
}

@ARTICLE{2025ApJ...990L..42C,
       author = {{Currie}, Thayne and {Hashimoto}, Jun and {Aoyama}, Yuhiko and {Dong}, Ruobing and {Fukagawa}, Misato and {Muto}, Takayuki and {Dykes}, Erica and {El Morsy}, Mona and {Tamura}, Motohide},
        title = "{VLT/MUSE Detection of the AB Aurigae b Protoplanet with H$_{{\ensuremath{\alpha}}}$ Spectroscopy}",
      journal = {\apjl},
         year = 2025,
        month = sep,
       volume = {990},
       number = {2},
          eid = {L42},
        pages = {L42},
          doi = {10.3847/2041-8213/adf7a0},
archivePrefix = {arXiv},
       eprint = {2508.18351},
 primaryClass = {astro-ph.EP},
       adsurl = {https://ui.adsabs.harvard.edu/abs/2025ApJ...990L..42C}
}

@ARTICLE{2022ApJ...941..172L,
       author = {{Leiendecker}, Harrison and {Jang-Condell}, Hannah and {Turner}, Neal J. and {Myers}, Adam D.},
        title = "{Dust Rings and Cavities in the Protoplanetary Disks around HD 163296 and DoAr 44}",
      journal = {\apj},
         year = 2022,
        month = dec,
       volume = {941},
       number = {2},
          eid = {172},
        pages = {172},
          doi = {10.3847/1538-4357/aca32d},
archivePrefix = {arXiv},
       eprint = {2212.00853},
 primaryClass = {astro-ph.EP},
       adsurl = {https://ui.adsabs.harvard.edu/abs/2022ApJ...941..172L}
}

@ARTICLE{2024A&A...682A..55M,
       author = {{Miley}, J.~M. and {Carpenter}, J. and {Booth}, R. and {Jennings}, J. and {Haworth}, T.~J. and {Vioque}, M. and {Andrews}, S. and {Wilner}, D. and {Benisty}, M. and {Huang}, J. and {Perez}, L. and {Guzman}, V. and {Ricci}, L. and {Isella}, A.},
        title = "{High-resolution ALMA observations of compact discs in the wide-binary system Sz 65 and Sz 66}",
      journal = {\aap},
         year = 2024,
        month = feb,
       volume = {682},
          eid = {A55},
        pages = {A55},
          doi = {10.1051/0004-6361/202347135},
archivePrefix = {arXiv},
       eprint = {2402.01903},
 primaryClass = {astro-ph.EP},
       adsurl = {https://ui.adsabs.harvard.edu/abs/2024A&A...682A..55M}
}

@ARTICLE{2025ApJ...989...69L,
       author = {{Lovell}, Joshua B. and {Trapman}, Leon and {Monsch}, Kristina and {Andrews}, Sean M. and {Booth}, Alice S. and {Keating}, Garrett K. and {Ueda}, Takahiro and {Wilner}, David J.},
        title = "{SMA and NOEMA Reveal Asymmetric Substructure in the Protoplanetary Disk of IRAS 23077+6707}",
      journal = {\apj},
         year = 2025,
        month = aug,
       volume = {989},
       number = {1},
          eid = {69},
        pages = {69},
          doi = {10.3847/1538-4357/ade392},
archivePrefix = {arXiv},
       eprint = {2507.00122},
 primaryClass = {astro-ph.EP},
       adsurl = {https://ui.adsabs.harvard.edu/abs/2025ApJ...989...69L}
}

@ARTICLE{2017A&A...601A...7F,
       author = {{Feldt}, M. and {Olofsson}, J. and {Boccaletti}, A. and {Maire}, A.~L. and {Milli}, J. and {Vigan}, A. and {Langlois}, M. and {Henning}, Th. and {Moor}, A. and {Bonnefoy}, M. and {Wahhaj}, Z. and {Desidera}, S. and {Gratton}, R. and {K{\'o}sp{\'a}l}, {\'A}. and {Abraham}, P. and {Menard}, F. and {Chauvin}, G. and {Lagrange}, A.~M. and {Mesa}, D. and {Salter}, G. and {Buenzli}, E. and {Lannier}, J. and {Perrot}, C. and {Peretti}, S. and {Sissa}, E.},
        title = "{SPHERE/SHINE reveals concentric rings in the debris disk of HIP 73145}",
      journal = {\aap},
         year = 2017,
        month = may,
       volume = {601},
          eid = {A7},
        pages = {A7},
          doi = {10.1051/0004-6361/201629261},
archivePrefix = {arXiv},
       eprint = {1612.07621},
 primaryClass = {astro-ph.SR},
       adsurl = {https://ui.adsabs.harvard.edu/abs/2017A&A...601A...7F}
}

@ARTICLE{2018A&A...610A..24F,
       author = {{Fedele}, D. and {Tazzari}, M. and {Booth}, R. and {Testi}, L. and {Clarke}, C.~J. and {Pascucci}, I. and {Kospal}, A. and {Semenov}, D. and {Bruderer}, S. and {Henning}, Th. and {Teague}, R.},
        title = "{ALMA continuum observations of the protoplanetary disk AS 209. Evidence of multiple gaps opened by a single planet}",
      journal = {\aap},
         year = 2018,
        month = feb,
       volume = {610},
          eid = {A24},
        pages = {A24},
          doi = {10.1051/0004-6361/201731978},
archivePrefix = {arXiv},
       eprint = {1711.05185},
 primaryClass = {astro-ph.EP},
       adsurl = {https://ui.adsabs.harvard.edu/abs/2018A&A...610A..24F}
}

@ARTICLE{2018ApJ...869L..49I,
       author = {{Isella}, Andrea and {Huang}, Jane and {Andrews}, Sean M. and {Dullemond}, Cornelis P. and {Birnstiel}, Tilman and {Zhang}, Shangjia and {Zhu}, Zhaohuan and {Guzm{\'a}n}, Viviana V. and {P{\'e}rez}, Laura M. and {Bai}, Xue-Ning and {Benisty}, Myriam and {Carpenter}, John M. and {Ricci}, Luca and {Wilner}, David J.},
        title = "{The Disk Substructures at High Angular Resolution Project (DSHARP). IX. A High-definition Study of the HD 163296 Planet-forming Disk}",
      journal = {\apjl},
         year = 2018,
        month = dec,
       volume = {869},
       number = {2},
          eid = {L49},
        pages = {L49},
          doi = {10.3847/2041-8213/aaf747},
archivePrefix = {arXiv},
       eprint = {1812.04047},
 primaryClass = {astro-ph.SR},
       adsurl = {https://ui.adsabs.harvard.edu/abs/2018ApJ...869L..49I}
}

@ARTICLE{2017AJ....154...33A,
       author = {{Avenhaus}, H. and {Quanz}, S.~P. and {Schmid}, H.~M. and {Dominik}, C. and {Stolker}, T. and {Ginski}, C. and {de Boer}, J. and {Szul{\'a}gyi}, J. and {Garufi}, A. and {Zurlo}, A. and {Hagelberg}, J. and {Benisty}, M. and {Henning}, T. and {M{\'e}nard}, F. and {Meyer}, M.~R. and {Baruffolo}, A. and {Bazzon}, A. and {Beuzit}, J.~L. and {Costille}, A. and {Dohlen}, K. and {Girard}, J.~H. and {Gisler}, D. and {Kasper}, M. and {Mouillet}, D. and {Pragt}, J. and {Roelfsema}, R. and {Salasnich}, B. and {Sauvage}, J.-F.},
        title = "{Exploring Dust around HD 142527 down to 0.″025 (4 au) Using SPHERE/ZIMPOL}",
      journal = {\aj},
         year = 2017,
        month = jul,
       volume = {154},
       number = {1},
          eid = {33},
        pages = {33},
          doi = {10.3847/1538-3881/aa7560},
archivePrefix = {arXiv},
       eprint = {1705.09680},
 primaryClass = {astro-ph.EP},
       adsurl = {https://ui.adsabs.harvard.edu/abs/2017AJ....154...33A}
}

@ARTICLE{2017A&A...607A..55V,
       author = {{van der Plas}, G. and {M{\'e}nard}, F. and {Canovas}, H. and {Avenhaus}, H. and {Casassus}, S. and {Pinte}, C. and {Caceres}, C. and {Cieza}, L.},
        title = "{An 80 au cavity in the disk around HD 34282}",
      journal = {\aap},
         year = 2017,
        month = nov,
       volume = {607},
          eid = {A55},
        pages = {A55},
          doi = {10.1051/0004-6361/201731392},
archivePrefix = {arXiv},
       eprint = {1707.01688},
 primaryClass = {astro-ph.SR},
       adsurl = {https://ui.adsabs.harvard.edu/abs/2017A&A...607A..55V}
}

@ARTICLE{2017A&A...597A..32V,
       author = {{van der Plas}, G. and {Wright}, C.~M. and {M{\'e}nard}, F. and {Casassus}, S. and {Canovas}, H. and {Pinte}, C. and {Maddison}, S.~T. and {Maaskant}, K. and {Avenhaus}, H. and {Cieza}, L. and {Perez}, S. and {Ubach}, C.},
        title = "{Cavity and other radial substructures in the disk around HD 97048}",
      journal = {\aap},
         year = 2017,
        month = jan,
       volume = {597},
          eid = {A32},
        pages = {A32},
          doi = {10.1051/0004-6361/201629523},
archivePrefix = {arXiv},
       eprint = {1609.02488},
 primaryClass = {astro-ph.SR},
       adsurl = {https://ui.adsabs.harvard.edu/abs/2017A&A...597A..32V}
}

@ARTICLE{2021MNRAS.502.5779N,
       author = {{Norfolk}, Brodie J. and {Maddison}, Sarah T. and {Pinte}, Christophe and {van der Marel}, Nienke and {Booth}, Richard A. and {Francis}, Logan and {Gonzalez}, Jean-Fran{\c{c}}ois and {M{\'e}nard}, Fran{\c{c}}ois and {Wright}, Chris M. and {van der Plas}, Gerrit and {Garg}, Himanshi},
        title = "{Dust traps and the formation of cavities in transition discs: a millimetre to sub-millimetre comparison survey}",
      journal = {\mnras},
         year = 2021,
        month = apr,
       volume = {502},
       number = {4},
        pages = {5779-5796},
          doi = {10.1093/mnras/stab313},
archivePrefix = {arXiv},
       eprint = {2102.02316},
 primaryClass = {astro-ph.EP},
       adsurl = {https://ui.adsabs.harvard.edu/abs/2021MNRAS.502.5779N}
}

@ARTICLE{2020A&A...637L...5B,
       author = {{Boccaletti}, A. and {Di Folco}, E. and {Pantin}, E. and {Dutrey}, A. and {Guilloteau}, S. and {Tang}, Y.~W. and {Pi{\'e}tu}, V. and {Habart}, E. and {Milli}, J. and {Beck}, T.~L. and {Maire}, A.-L.},
        title = "{Possible evidence of ongoing planet formation in AB Aurigae. A showcase of the SPHERE/ALMA synergy}",
      journal = {\aap},
         year = 2020,
        month = may,
       volume = {637},
          eid = {L5},
        pages = {L5},
          doi = {10.1051/0004-6361/202038008},
archivePrefix = {arXiv},
       eprint = {2005.09064},
 primaryClass = {astro-ph.EP},
       adsurl = {https://ui.adsabs.harvard.edu/abs/2020A&A...637L...5B}
}

@ARTICLE{2021A&A...654A..35B,
       author = {{Brown-Sevilla}, S.~B. and {Keppler}, M. and {Barraza-Alfaro}, M. and {Melon Fuksman}, J.~D. and {Kurtovic}, N. and {Pinilla}, P. and {Feldt}, M. and {Brandner}, W. and {Ginski}, C. and {Henning}, Th. and {Klahr}, H. and {Asensio-Torres}, R. and {Cantalloube}, F. and {Garufi}, A. and {van Holstein}, R.~G. and {Langlois}, M. and {M{\'e}nard}, F. and {Rickman}, E. and {Benisty}, M. and {Chauvin}, G. and {Zurlo}, A. and {Weber}, P. and {Pavlov}, A. and {Ramos}, J. and {Rochat}, S. and {Roelfsema}, R.},
        title = "{A multiwavelength analysis of the spiral arms in the protoplanetary disk around WaOph 6}",
      journal = {\aap},
         year = 2021,
        month = oct,
       volume = {654},
          eid = {A35},
        pages = {A35},
          doi = {10.1051/0004-6361/202140783},
archivePrefix = {arXiv},
       eprint = {2107.13560},
 primaryClass = {astro-ph.EP},
       adsurl = {https://ui.adsabs.harvard.edu/abs/2021A&A...654A..35B}
}

@ARTICLE{2025A&A...699A.237G,
       author = {{Ginski}, C. and {Pinilla}, P. and {Benisty}, M. and {Pinte}, C. and {Claes}, R. and {Mamajek}, E. and {Kenworthy}, M. and {Murphy}, M. and {Manara}, C. and {Bae}, J. and {Birnstiel}, T. and {Byrne}, J. and {Dominik}, C. and {Facchini}, S. and {Garufi}, A. and {Gratton}, R. and {Hogerheijde}, M. and {van Holstein}, R. and {Huang}, J. and {Langlois}, M. and {Lawlor}, C. and {Ma}, J. and {McLachlan}, D. and {Menard}, F. and {Rigliaco}, R. and {Ribas}, A. and {Schmidt}, T. and {Sierra}, A. and {Tazaki}, R. and {Williams}, J. and {Zurlo}, A.},
        title = "{Disk Evolution Study Through Imaging of Nearby Young Stars (DESTINYS): Evidence of planet{\textendash}disk interaction in the 2MASSJ16120668-3010270 system}",
      journal = {\aap},
         year = 2025,
        month = jul,
       volume = {699},
          eid = {A237},
        pages = {A237},
          doi = {10.1051/0004-6361/202451647},
archivePrefix = {arXiv},
       eprint = {2506.05892},
 primaryClass = {astro-ph.EP},
       adsurl = {https://ui.adsabs.harvard.edu/abs/2025A&A...699A.237G}
}

@INPROCEEDINGS{pedichini22,
       author = {{Pedichini}, Fernando and {Piazzesi}, Roberto and {Viavattene}, Giorgio and {Antoniucci}, Simone and {Gangi}, Manuele and {Li Causi}, Gianluca and {Mattioli}, Massimiliano and {Stangalini}, Marco and {Terreri}, Alessandro and {Testa}, Vincenzo and {Pinna}, Enrico and {Briguglio}, Runa and {Agapito}, Guido and {Puglisi}, Alfio and {Farinato}, Jacopo and {Bergomi}, Maria},
        title = "{SHARK-VIS ready for the stars: instrument description and final laboratory performance test}",
    booktitle = {Adaptive Optics Systems VIII},
         year = 2022,
       editor = {{Schreiber}, Laura and {Schmidt}, Dirk and {Vernet}, Elise},
       series = {Society of Photo-Optical Instrumentation Engineers (SPIE) Conference Series},
       volume = {12185},
        month = aug,
          eid = {121856Q},
        pages = {121856Q},
          doi = {10.1117/12.2629244},
       adsurl = {https://ui.adsabs.harvard.edu/abs/2022SPIE12185E..6QP}
}

@INPROCEEDINGS{pedichini24,
       author = {{Pedichini}, Fernando and {Piazzesi}, Roberto and {Li Causi}, Gianluca and {Antoniucci}, Simone and {Testa}, Vincenzo and {Vaccari}, Piero and {Farinato}, Jacopo and {Bergomi}, Maria and {Mesa}, Dino and {D'Orazi}, Valentina and {Esposito}, Simone and {Pinna}, Enrico and {Cardwell}, Andrew and {Guerra}, Juan Carlos and {Power}, Jennifer and {Ertel}, Steve and {Montoya}, Manny and {Carlson}, Jared},
        title = "{Early results from AO assisted visible observations with the new high-contrast imager SHARK-VIS at the Large Binocular Telescope}",
    booktitle = {Adaptive Optics Systems IX},
         year = 2024,
       editor = {{Jackson}, Kathryn J. and {Schmidt}, Dirk and {Vernet}, Elise},
       series = {Society of Photo-Optical Instrumentation Engineers (SPIE) Conference Series},
       volume = {13097},
        month = aug,
          eid = {130970A},
        pages = {130970A},
          doi = {10.1117/12.3018651},
       adsurl = {https://ui.adsabs.harvard.edu/abs/2024SPIE13097E..0AP}
}

@ARTICLE{Li2025,
       author = {{Li}, Jialin and {Close}, Laird M. and {Long}, Feng and {Males}, Jared R. and {Haffert}, Sebastiaan Y. and {Weinberger}, Alycia and {Follette}, Katherine and {Andrews}, Sean and {Carpenter}, John and {Foster}, Warren B. and {Van Gorkom}, Kyle and {Hedglen}, Alexander D. and {Herczeg}, Gregory J. and {Johnson}, Parker T. and {Kautz}, Maggie Y. and {Kueny}, Jay K. and {Li}, Rixin and {Liberman}, Joshua and {Long}, Joseph D. and {Lumbres}, Jennifer and {Marino}, Sebastian and {Matr{\`a}}, Luca and {McEwen}, Eden A. and {Guyon}, Olivier and {Pearce}, Logan A. and {P{\'e}rez}, Laura M. and {Pinilla}, Paola and {Schatz}, Lauren and {Shi}, Yangfan and {Twitchell}, Katie and {Wagner}, Kevin and {Wilner}, David and {Wu}, Ya-Lin and {Zhang}, Shangjia and {Zhu}, Zhaohuan},
        title = "{Discovery of H{\ensuremath{\alpha}} Emission from a Protoplanet Candidate around the Young Star 2MASS J16120668─3010270 with MagAO-X}",
      journal = {\apjl},
         year = 2025,
        month = sep,
       volume = {990},
       number = {2},
          eid = {L70},
        pages = {L70},
          doi = {10.3847/2041-8213/adfcbd},
archivePrefix = {arXiv},
       eprint = {2508.11155},
 primaryClass = {astro-ph.EP},
       adsurl = {https://ui.adsabs.harvard.edu/abs/2025ApJ...990L..70L}
}

@ARTICLE{Close2025,
       author = {{Close}, Laird M. and {van Capelleveen}, Richelle F. and {Weible}, Gabriel and {Wagner}, Kevin and {Haffert}, Sebastiaan Y. and {Males}, Jared R. and {Ilyin}, Ilya and {Kenworthy}, Matthew A. and {Li}, Jialin and {Long}, Joseph D. and {Ertel}, Steve and {Ginski}, Christian and {Weinberger}, Alycia J. and {Follette}, Kate and {Liberman}, Joshua and {Twitchell}, Katie and {Johnson}, Parker and {Kueny}, Jay and {Apai}, Daniel and {Doyon}, Rene and {Foster}, Warren and {Gasho}, Victor and {Van Gorkom}, Kyle and {Guyon}, Olivier and {Kautz}, Maggie Y. and {McLeod}, Avalon and {McEwen}, Eden and {Pearce}, Logan and {Schatz}, Lauren and {Hedglen}, Alexander D. and {Wu}, Ya-Lin and {Isbell}, Jacob and {Power}, Jenny and {Carlson}, Jared and {Close}, Emmeline and {Tonucci}, Elena and {Mars}, Matthijs},
        title = "{Wide Separation Planets in Time (WISPIT): Discovery of a Gap H{\ensuremath{\alpha}} Protoplanet WISPIT 2b with MagAO-X}",
      journal = {\apjl},
         year = 2025,
        month = sep,
       volume = {990},
       number = {1},
          eid = {L9},
        pages = {L9},
          doi = {10.3847/2041-8213/adf7a5},
archivePrefix = {arXiv},
       eprint = {2508.19046},
 primaryClass = {astro-ph.EP},
       adsurl = {https://ui.adsabs.harvard.edu/abs/2025ApJ...990L...9C}
}

@ARTICLE{Bowler2025,
       author = {{Bowler}, Brendan P. and {Zhou}, Yifan and {Biddle}, Lauren I. and {Jiang}, Lillian Yushu and {Bae}, Jaehan and {Close}, Laird M. and {Follette}, Katherine B. and {Franson}, Kyle and {Kraus}, Adam L. and {Sanghi}, Aniket and {Tran}, Quang and {Ward-Duong}, Kimberly and {Wu}, Ya-Lin and {Zhu}, Zhaohuan},
        title = "{H{\ensuremath{\alpha}} Variability of AB Aur b with the Hubble Space Telescope: Probing the Nature of a Protoplanet Candidate with Accretion Light Echoes}",
      journal = {\aj},
         year = 2025,
        month = may,
       volume = {169},
       number = {5},
          eid = {258},
        pages = {258},
          doi = {10.3847/1538-3881/adb6a1},
archivePrefix = {arXiv},
       eprint = {2502.14736},
 primaryClass = {astro-ph.EP},
       adsurl = {https://ui.adsabs.harvard.edu/abs/2025AJ....169..258B}
}

@ARTICLE{Hammond2023,
       author = {{Hammond}, Iain and {Christiaens}, Valentin and {Price}, Daniel J. and {Toci}, Claudia and {Pinte}, Christophe and {Juillard}, Sandrine and {Garg}, Himanshi},
        title = "{Confirmation and Keplerian motion of the gap-carving protoplanet HD 169142 b}",
      journal = {\mnras},
         year = 2023,
        month = jun,
       volume = {522},
       number = {1},
        pages = {L51-L55},
          doi = {10.1093/mnrasl/slad027},
archivePrefix = {arXiv},
       eprint = {2302.11302},
 primaryClass = {astro-ph.EP},
       adsurl = {https://ui.adsabs.harvard.edu/abs/2023MNRAS.522L..51H}
}

@ARTICLE{Zhu2015,
       author = {{Zhu}, Zhaohuan and {Dong}, Ruobing and {Stone}, James M. and {Rafikov}, Roman R.},
        title = "{The Structure of Spiral Shocks Excited by Planetary-mass Companions}",
      journal = {\apj},
         year = 2015,
        month = nov,
       volume = {813},
       number = {2},
          eid = {88},
        pages = {88},
          doi = {10.1088/0004-637X/813/2/88},
archivePrefix = {arXiv},
       eprint = {1507.03599},
 primaryClass = {astro-ph.SR},
       adsurl = {https://ui.adsabs.harvard.edu/abs/2015ApJ...813...88Z}
}

@ARTICLE{Kanagawa2016,
       author = {{Kanagawa}, Kazuhiro D. and {Muto}, Takayuki and {Tanaka}, Hidekazu and {Tanigawa}, Takayuki and {Takeuchi}, Taku and {Tsukagoshi}, Takashi and {Momose}, Munetake},
        title = "{Mass constraint for a planet in a protoplanetary disk from the gap width}",
      journal = {\pasj},
         year = 2016,
        month = jun,
       volume = {68},
       number = {3},
          eid = {43},
        pages = {43},
          doi = {10.1093/pasj/psw037},
archivePrefix = {arXiv},
       eprint = {1603.03853},
 primaryClass = {astro-ph.EP},
       adsurl = {https://ui.adsabs.harvard.edu/abs/2016PASJ...68...43K}
}

@ARTICLE{Fung2015,
       author = {{Fung}, Jeffrey and {Dong}, Ruobing},
        title = "{Inferring Planet Mass from Spiral Structures in Protoplanetary Disks}",
      journal = {\apjl},
         year = 2015,
        month = dec,
       volume = {815},
       number = {2},
          eid = {L21},
        pages = {L21},
          doi = {10.1088/2041-8205/815/2/L21},
archivePrefix = {arXiv},
       eprint = {1511.01178},
 primaryClass = {astro-ph.EP},
       adsurl = {https://ui.adsabs.harvard.edu/abs/2015ApJ...815L..21F}
}

@ARTICLE{Baraffe1998,
       author = {{Baraffe}, I. and {Chabrier}, G. and {Allard}, F. and {Hauschildt}, P.~H.},
        title = "{Evolutionary models for solar metallicity low-mass stars: mass-magnitude relationships and color-magnitude diagrams}",
      journal = {\aap},
         year = 1998,
        month = sep,
       volume = {337},
        pages = {403-412},
          doi = {10.48550/arXiv.astro-ph/9805009},
archivePrefix = {arXiv},
       eprint = {astro-ph/9805009},
 primaryClass = {astro-ph},
       adsurl = {https://ui.adsabs.harvard.edu/abs/1998A&A...337..403B}
}

@ARTICLE{Chabrier2000,
       author = {{Chabrier}, G. and {Baraffe}, I. and {Allard}, F. and {Hauschildt}, P.},
        title = "{Evolutionary Models for Very Low-Mass Stars and Brown Dwarfs with Dusty Atmospheres}",
      journal = {\apj},
         year = 2000,
        month = oct,
       volume = {542},
       number = {1},
        pages = {464-472},
          doi = {10.1086/309513},
archivePrefix = {arXiv},
       eprint = {astro-ph/0005557},
 primaryClass = {astro-ph},
       adsurl = {https://ui.adsabs.harvard.edu/abs/2000ApJ...542..464C}
}

@ARTICLE{2025ApJ...990L...8V,
       author = {{van Capelleveen}, Richelle F. and {Ginski}, Christian and {Kenworthy}, Matthew A. and {Byrne}, Jake and {Lawlor}, Chloe and {McLachlan}, Dan and {Mamajek}, Eric E. and {Stolker}, Tomas and {Benisty}, Myriam and {Bohn}, Alexander J. and {Close}, Laird M. and {Dominik}, Carsten and {Haffert}, Sebastiaan and {Landman}, Rico and {Ma}, Jie and {Snellen}, Ignas and {Tazaki}, Ryo and {van der Marel}, Nienke and {Welzel}, Lukas and {Zhang}, Yapeng},
        title = "{WIde Separation Planets In Time (WISPIT): A Gap-clearing Planet in a Multi-ringed Disk around the Young Solar-type Star WISPIT 2}",
      journal = {\apjl},
         year = 2025,
        month = sep,
       volume = {990},
       number = {1},
          eid = {L8},
        pages = {L8},
          doi = {10.3847/2041-8213/adf721},
archivePrefix = {arXiv},
       eprint = {2508.19053},
 primaryClass = {astro-ph.EP},
       adsurl = {https://ui.adsabs.harvard.edu/abs/2025ApJ...990L...8V}
}

@ARTICLE{vanderMarel2019,
       author = {{van der Marel}, Nienke and {Dong}, Ruobing and {di Francesco}, James and {Williams}, Jonathan P. and {Tobin}, John},
        title = "{Protoplanetary Disk Rings and Gaps across Ages and Luminosities}",
      journal = {\apj},
         year = 2019,
        month = feb,
       volume = {872},
       number = {1},
          eid = {112},
        pages = {112},
          doi = {10.3847/1538-4357/aafd31},
archivePrefix = {arXiv},
       eprint = {1901.03680},
 primaryClass = {astro-ph.EP},
       adsurl = {https://ui.adsabs.harvard.edu/abs/2019ApJ...872..112V}
}

@ARTICLE{Rafikov2011,
       author = {{Rafikov}, Roman R.},
        title = "{Constraint on the Giant Planet Production by Core Accretion}",
      journal = {\apj},
         year = 2011,
        month = feb,
       volume = {727},
       number = {2},
          eid = {86},
        pages = {86},
          doi = {10.1088/0004-637X/727/2/86},
archivePrefix = {arXiv},
       eprint = {1004.5139},
 primaryClass = {astro-ph.EP},
       adsurl = {https://ui.adsabs.harvard.edu/abs/2011ApJ...727...86R}
}

@INPROCEEDINGS{Helled2014,
       author = {{Helled}, R. and {Bodenheimer}, P. and {Podolak}, M. and {Boley}, A. and {Meru}, F. and {Nayakshin}, S. and {Fortney}, J.~J. and {Mayer}, L. and {Alibert}, Y. and {Boss}, A.~P.},
        title = "{Giant Planet Formation, Evolution, and Internal Structure}",
    booktitle = {Protostars and Planets VI},
         year = 2014,
       editor = {{Beuther}, Henrik and {Klessen}, Ralf S. and {Dullemond}, Cornelis P. and {Henning}, Thomas},
        month = jan,
        pages = {643-665},
          doi = {10.2458/azu_uapress_9780816531240-ch028},
archivePrefix = {arXiv},
       eprint = {1311.1142},
 primaryClass = {astro-ph.EP},
       adsurl = {https://ui.adsabs.harvard.edu/abs/2014prpl.conf..643H}
}

@ARTICLE{Gurrutxaga2024,
       author = {{Gurrutxaga}, Nerea and {Johansen}, Anders and {Lambrechts}, Michiel and {Appelgren}, Johan},
        title = "{Formation of wide-orbit giant planets in protoplanetary disks with a decreasing pebble flux}",
      journal = {\aap},
         year = 2024,
        month = feb,
       volume = {682},
          eid = {A43},
        pages = {A43},
          doi = {10.1051/0004-6361/202348020},
archivePrefix = {arXiv},
       eprint = {2311.04365},
 primaryClass = {astro-ph.EP},
       adsurl = {https://ui.adsabs.harvard.edu/abs/2024A&A...682A..43G}
}

@ARTICLE{Feiden2016,
       author = {{Feiden}, Gregory A.},
        title = "{Magnetic inhibition of convection and the fundamental properties of low-mass stars. III. A consistent 10 Myr age for the Upper Scorpius OB association}",
      journal = {\aap},
         year = 2016,
        month = sep,
       volume = {593},
          eid = {A99},
        pages = {A99},
          doi = {10.1051/0004-6361/201527613},
archivePrefix = {arXiv},
       eprint = {1604.08036},
 primaryClass = {astro-ph.SR},
       adsurl = {https://ui.adsabs.harvard.edu/abs/2016A&A...593A..99F}
}

@ARTICLE{Lawlor2026,
       author = {{Lawlor}, Chloe and {van Capelleveen}, Richelle F. and {Bourdarot}, Guillaume and {Ginski}, Christian and {Kenworthy}, Matthew A. and {Stolker}, Tomas and {Close}, Laird and {Bohn}, Alexander J. and {Eisenhauer}, Frank and {Garcia}, Paulo and {H{\"o}nig}, Sebastian F. and {Kammerer}, Jens and {Kreidberg}, Laura and {Lacour}, Sylvestre and {Le Bouquin}, Jean-Baptiste and {Mamajek}, Eric and {Nowak}, Mathias and {Paumard}, Thibaut and {Straubmeier}, Christian and {van der Marel}, Nienke and {The Exogravity Collaboration}},
        title = "{Direct Spectroscopic Confirmation of the Young Embedded Protoplanet WISPIT 2c}",
      journal = {\apjl},
         year = 2026,
        month = apr,
       volume = {1000},
       number = {2},
          eid = {L38},
        pages = {L38},
          doi = {10.3847/2041-8213/ae4b3b},
archivePrefix = {arXiv},
       eprint = {2603.22085},
 primaryClass = {astro-ph.EP},
       adsurl = {https://ui.adsabs.harvard.edu/abs/2026ApJ..1000L..38L}
}

@ARTICLE{Facchini2026,
       author = {{Facchini}, Stefano and {Curone}, Pietro and {Benisty}, Myriam and {Zagaria}, Francesco and {Teague}, Richard and {Cugno}, Gabriele and {Bae}, Jaehan},
        title = "{A 2 au Resolution View by ALMA of the Planet-hosting WISPIT 2 Disk}",
      journal = {\apjl},
         year = 2026,
        month = feb,
       volume = {998},
       number = {1},
          eid = {L16},
        pages = {L16},
          doi = {10.3847/2041-8213/ae3c0a},
archivePrefix = {arXiv},
       eprint = {2601.15948},
 primaryClass = {astro-ph.EP},
       adsurl = {https://ui.adsabs.harvard.edu/abs/2026ApJ...998L..16F}
}

@ARTICLE{1998ApJ...503..923B,
       author = {{Boss}, Alan P.},
        title = "{Evolution of the Solar Nebula. IV. Giant Gaseous Protoplanet Formation}",
      journal = {\apj},
         year = 1998,
        month = aug,
       volume = {503},
       number = {2},
        pages = {923-937},
          doi = {10.1086/306036},
       adsurl = {https://ui.adsabs.harvard.edu/abs/1998ApJ...503..923B}
}

@ARTICLE{2013A&A...552A.129V,
       author = {{Vorobyov}, E.~I.},
        title = "{Formation of giant planets and brown dwarfs on wide orbits}",
      journal = {\aap},
         year = 2013,
        month = apr,
       volume = {552},
          eid = {A129},
        pages = {A129},
          doi = {10.1051/0004-6361/201220601},
archivePrefix = {arXiv},
       eprint = {1302.1892},
 primaryClass = {astro-ph.EP},
       adsurl = {https://ui.adsabs.harvard.edu/abs/2013A&A...552A.129V}
}

@ARTICLE{2022A&A...668A..50S,
       author = {{Stapper}, L.~M. and {Ginski}, C.},
        title = "{Iterative angular differential imaging (IADI): An exploration of recovering disk structures in scattered light with an iterative ADI approach}",
      journal = {\aap},
         year = 2022,
        month = dec,
       volume = {668},
          eid = {A50},
        pages = {A50},
          doi = {10.1051/0004-6361/202142820},
archivePrefix = {arXiv},
       eprint = {2210.02454},
 primaryClass = {astro-ph.IM},
       adsurl = {https://ui.adsabs.harvard.edu/abs/2022A&A...668A..50S}
}

@ARTICLE{2009ApJ...693.1056L,
       author = {{Lorenzetti}, D. and {Larionov}, V.~M. and {Giannini}, T. and {Arkharov}, A.~A. and {Antoniucci}, S. and {Nisini}, B. and {Di Paola}, A.},
        title = "{Near-Infrared Spectroscopic Monitoring of EXor Variables: First Results}",
      journal = {\apj},
         year = 2009,
        month = mar,
       volume = {693},
       number = {2},
        pages = {1056-1073},
          doi = {10.1088/0004-637X/693/2/1056},
archivePrefix = {arXiv},
       eprint = {0811.1164},
 primaryClass = {astro-ph},
       adsurl = {https://ui.adsabs.harvard.edu/abs/2009ApJ...693.1056L}
}

@INPROCEEDINGS{2016SPIE.9909E..3VP,
       author = {{Pinna}, E. and {Esposito}, S. and {Hinz}, P. and {Agapito}, G. and {Bonaglia}, M. and {Puglisi}, A. and {Xompero}, M. and {Riccardi}, A. and {Briguglio}, R. and {Arcidiacono}, C. and {Carbonaro}, L. and {Fini}, L. and {Montoya}, M. and {Durney}, O.},
        title = "{SOUL: the Single conjugated adaptive Optics Upgrade for LBT}",
    booktitle = {Adaptive Optics Systems V},
         year = 2016,
       editor = {{Marchetti}, Enrico and {Close}, Laird M. and {V{\'e}ran}, Jean-Pierre},
       series = {Society of Photo-Optical Instrumentation Engineers (SPIE) Conference Series},
       volume = {9909},
        month = jul,
          eid = {99093V},
        pages = {99093V},
          doi = {10.1117/12.2234444},
       adsurl = {https://ui.adsabs.harvard.edu/abs/2016SPIE.9909E..3VP}
}

@INPROCEEDINGS{2023aoel.confE..80P,
       author = {{Pinna}, Enrico and {Rossi}, Fabio and {Agapito}, Guido and {Puglisi}, Alfio and {Plantet}, Cedric and {Ghose}, Essna and {Matthieu}, Bec and {Bonaglia}, Marco and {Briguglio}, Runa and {Brusa}, Guido and {Carbonaro}, Luca and {Cavallaro}, Alessandro and {Christou}, Julian and {Durney}, Olivier and {Ertel}, Steve and {Esposito}, Simone and {Grani}, Paolo and {Hinz}, Philip and {Guerra}, Juan Carlos and {Lefebvre}, Michael and {Mazzoni}, Tommaso and {Mechtley}, Brandon and {Miller}, Douglas L. and {Montoya}, Manny and {Power}, Jennifer and {Rothberg}, Barry and {Taylor}, Gregory and {Vaz}, Amali and {Xompero}, Marco and {Zhang}, Xianyu},
        title = "{SOUL at LBT: commissioning results, science and future.}",
    booktitle = {Adaptive Optics for Extremely Large Telescopes (AO4ELT7)},
         year = 2023,
        month = jun,
          eid = {80},
        pages = {80},
          doi = {10.13009/AO4ELT7-2023-082},
archivePrefix = {arXiv},
       eprint = {2310.14447},
 primaryClass = {astro-ph.IM},
       adsurl = {https://ui.adsabs.harvard.edu/abs/2023aoel.confE..80P}
}

@ARTICLE{2017AJ....153..264F,
       author = {{Follette}, Katherine B. and {Rameau}, Julien and {Dong}, Ruobing and {Pueyo}, Laurent and {Close}, Laird M. and {Duch{\^e}ne}, Gaspard and {Fung}, Jeffrey and {Leonard}, Clare and {Macintosh}, Bruce and {Males}, Jared R. and {Marois}, Christian and {Millar-Blanchaer}, Maxwell A. and {Morzinski}, Katie M. and {Mullen}, Wyatt and {Perrin}, Marshall and {Spiro}, Elijah and {Wang}, Jason and {Ammons}, S. Mark and {Bailey}, Vanessa P. and {Barman}, Travis and {Bulger}, Joanna and {Chilcote}, Jeffrey and {Cotten}, Tara and {De Rosa}, Robert J. and {Doyon}, Rene and {Fitzgerald}, Michael P. and {Goodsell}, Stephen J. and {Graham}, James R. and {Greenbaum}, Alexandra Z. and {Hibon}, Pascale and {Hung}, Li-Wei and {Ingraham}, Patrick and {Kalas}, Paul and {Konopacky}, Quinn and {Larkin}, James E. and {Maire}, J{\'e}r{\^o}me and {Marchis}, Franck and {Metchev}, Stanimir and {Nielsen}, Eric L. and {Oppenheimer}, Rebecca and {Palmer}, David and {Patience}, Jennifer and {Poyneer}, Lisa and {Rajan}, Abhijith and {Rantakyr{\"o}}, Fredrik T. and {Savransky}, Dmitry and {Schneider}, Adam C. and {Sivaramakrishnan}, Anand and {Song}, Inseok and {Soummer}, Remi and {Thomas}, Sandrine and {Vega}, David and {Wallace}, J. Kent and {Ward-Duong}, Kimberly and {Wiktorowicz}, Sloane and {Wolff}, Schuyler},
        title = "{Complex Spiral Structure in the HD 100546 Transitional Disk as Revealed by GPI and MagAO}",
      journal = {\aj},
         year = 2017,
        month = jun,
       volume = {153},
       number = {6},
          eid = {264},
        pages = {264},
          doi = {10.3847/1538-3881/aa6d85},
archivePrefix = {arXiv},
       eprint = {1704.06260},
 primaryClass = {astro-ph.EP},
       adsurl = {https://ui.adsabs.harvard.edu/abs/2017AJ....153..264F}
}

@ARTICLE{2015ApJ...812L..32D,
       author = {{Dong}, Ruobing and {Hall}, Cassandra and {Rice}, Ken and {Chiang}, Eugene},
        title = "{Spiral Arms in Gravitationally Unstable Protoplanetary Disks as Imaged in Scattered Light}",
      journal = {\apjl},
         year = 2015,
        month = oct,
       volume = {812},
       number = {2},
          eid = {L32},
        pages = {L32},
          doi = {10.1088/2041-8205/812/2/L32},
archivePrefix = {arXiv},
       eprint = {1510.00396},
 primaryClass = {astro-ph.SR},
       adsurl = {https://ui.adsabs.harvard.edu/abs/2015ApJ...812L..32D}
}

@ARTICLE{1964ApJ...139.1217T,
       author = {{Toomre}, A.},
        title = "{On the gravitational stability of a disk of stars.}",
      journal = {\apj},
         year = 1964,
        month = may,
       volume = {139},
        pages = {1217-1238},
          doi = {10.1086/147861},
       adsurl = {https://ui.adsabs.harvard.edu/abs/1964ApJ...139.1217T}
}

@ARTICLE{2014MNRAS.441.2094R,
       author = {{Rosotti}, Giovanni P. and {Dale}, James E. and {de Juan Ovelar}, Maria and {Hubber}, David A. and {Kruijssen}, J.~M. Diederik and {Ercolano}, Barbara and {Walch}, Stefanie},
        title = "{Protoplanetary disc evolution affected by star-disc interactions in young stellar clusters}",
      journal = {\mnras},
         year = 2014,
        month = jul,
       volume = {441},
       number = {3},
        pages = {2094-2110},
          doi = {10.1093/mnras/stu679},
archivePrefix = {arXiv},
       eprint = {1404.1931},
 primaryClass = {astro-ph.EP},
       adsurl = {https://ui.adsabs.harvard.edu/abs/2014MNRAS.441.2094R}
}

@ARTICLE{2019A&A...628A..20D,
       author = {{Dullemond}, C.~P. and {K{\"u}ffmeier}, M. and {Goicovic}, F. and {Fukagawa}, M. and {Oehl}, V. and {Kramer}, M.},
        title = "{Cloudlet capture by transitional disk and FU Orionis stars}",
      journal = {\aap},
         year = 2019,
        month = aug,
       volume = {628},
          eid = {A20},
        pages = {A20},
          doi = {10.1051/0004-6361/201832632},
archivePrefix = {arXiv},
       eprint = {1911.05158},
 primaryClass = {astro-ph.EP},
       adsurl = {https://ui.adsabs.harvard.edu/abs/2019A&A...628A..20D}
}

@ARTICLE{2020A&A...633A...3K,
       author = {{Kuffmeier}, M. and {Goicovic}, F.~G. and {Dullemond}, C.~P.},
        title = "{Late encounter events as source of disks and spiral structures. Forming second generation disks}",
      journal = {\aap},
         year = 2020,
        month = jan,
       volume = {633},
          eid = {A3},
        pages = {A3},
          doi = {10.1051/0004-6361/201936820},
archivePrefix = {arXiv},
       eprint = {1911.04833},
 primaryClass = {astro-ph.SR},
       adsurl = {https://ui.adsabs.harvard.edu/abs/2020A&A...633A...3K}
}

@ARTICLE{2025MNRAS.537.2695C,
       author = {{Calcino}, Josh and {Price}, Daniel J. and {Hilder}, Thomas and {Christiaens}, Valentin and {Speedie}, Jessica and {Ormel}, Chris W.},
        title = "{Anatomy of a fall: stationary and super-Keplerian spiral arms generated by accretion streamers in protostellar discs}",
      journal = {\mnras},
         year = 2025,
        month = mar,
       volume = {537},
       number = {3},
        pages = {2695-2707},
          doi = {10.1093/mnras/staf135},
archivePrefix = {arXiv},
       eprint = {2410.18521},
 primaryClass = {astro-ph.EP},
       adsurl = {https://ui.adsabs.harvard.edu/abs/2025MNRAS.537.2695C}
}

@ARTICLE{2007A&A...469..963P,
       author = {{Pinte}, C. and {Fouchet}, L. and {M{\'e}nard}, F. and {Gonzalez}, J.-F. and {Duch{\^e}ne}, G.},
        title = "{On the stratified dust distribution of the GG Tauri circumbinary ring}",
      journal = {\aap},
         year = 2007,
        month = jul,
       volume = {469},
       number = {3},
        pages = {963-971},
          doi = {10.1051/0004-6361:20077137},
archivePrefix = {arXiv},
       eprint = {0704.2747},
 primaryClass = {astro-ph},
       adsurl = {https://ui.adsabs.harvard.edu/abs/2007A&A...469..963P}
}

\end{document}